\documentclass[reprint,secnumarabic,amssymb, nobibnotes, nofootinbib, aps, prd, onecolumn]{revtex4-2}
\usepackage{graphicx}
\usepackage{amsmath}
\usepackage{epstopdf}
\usepackage{setspace}

\begin{document}

\title{Non-coherent evolution of closed weakly interacting system leads to \\ equidistribution of probabilities of microstates}

\author{A.\,P. Meilakhs} 
\email[A.\,P. Meilakhs: ]{iestfi@gmail.com}
\affiliation{$^1$Departamento de Fisica de Materia Condensada, GIYA, CAC-CNEA, Av. Gral. Paz 1499, San Martin, Pcia. Buenos Aires, Argentina \\
 $^2$Instituto de Nanociencia y Nanotecnologia, INN-CONICET-CNEA}
\date{\today}

\begin{abstract}

We introduce a concept of non-coherent evolution of macroscopic quantum systems. We show that for weakly interacting systems such evolution is a Markovian stochastic process. The transition rates between system states, which characterize the process, are determined by Fermi's golden rule. Such evolution is time-irreversible and leads to the equidistribution of probabilities across every state of the system. Furthermore, we investigate the time dependence of the mean numbers of particles in single-particle states and find that, under the given assumptions, it is governed by the Boltzmann collision integral. The proposed mechanism that transforms time-reversible unitary evolution into time-irreversible stochastic evolution is non-coherence. In the presented theory, the non-coherence is not associated with interaction with a heat bath, but rather with the finite spectral width of quantum states. This understanding of non-coherence is analogous to the one used in wave optics. Thus, we present a novel approach to the famous arrow of time problem.
\end{abstract}

\maketitle

\section{Introduction}

The arrow-of-time problem, as presented in textbooks on statistical mechanics \cite{LdStat, Prig} and in specialized books on the nature of time \cite{Time1, Time2}, remains one of the most fundamental challenges in modern physics. The issue arises from the time asymmetry in the equations governing the behavior of large ensembles of particles, in contrast to the time-symmetric nature of the equations describing individual particles.

Let us list the well-known time-asymmetric properties of macroscopic systems. Probably the most famous is the second law of thermodynamics, which states that the entropy of an isolated system undergoing spontaneous evolution cannot decrease \cite{Prig}. Next is the existence of thermodynamic equilibrium—a state toward which an isolated system evolves. This state is characterized by the equidistribution of probabilities among microstates. The state in which all microstates are equally probable is called the Gibbs microcanonical ensemble \cite{Kubo}. Finally, the Boltzmann equation, which describes the time evolution of single-particle distribution functions, is famously irreversible.

The classical approach to deriving the Boltzmann equation is the BBGKY hierarchy \cite{BBGKY}. However, it postulates molecular chaos as a model for the two-particle distribution function without fully elucidating its origins. Another interesting attempt to deduce the Boltzmann equation is found in Refs. \cite{MeanField1, MeanField2}, where it is treated as a mean-field theory; however, only the diffusive, reversible part can be derived in this way. In Ref. \cite{Gibbs}, the authors take the reverse route -- not deriving the concept of probability equidistribution, but rather arguing that statistical physics can be formulated without this assumption.

Other approaches to explaining asymmetry postulate low-entropy initial states of the universe \cite{Universe, NofTime}, which is not completely satisfactory, as it constitutes an ad hoc hypothesis \cite{AdHoc}. In contrast, Ref. \cite{Modify} proposes that physical laws themselves must be modified to incorporate irreversibility.

Modern approaches propose that irreversibility emerges from a time-asymmetric mechanism inherent in quantum mechanics \cite{Quantum}. Decoherence is typically suggested as this mechanism \cite{DecohTime}. Decoherence theory explains measurement in quantum theory as the interaction of a quantum system with its environment, leading to the loss of coherence in the system \cite{DecohBook}. However, it is not universally accepted that decoherence theory resolves the measurement problem \cite{DecohProblem1, DecohProblem2, DecohProblem3}. While it does account for the vanishing of the non-diagonal elements of the density matrix, it does not explain why only one definite outcome is realized in a measurement. We will refer to this issue as the problem of outcomes.

Another notable approach is the study of quantum entanglement and its relationship to thermodynamic irreversibility. Some researchers have proposed that the generation and disruption of entanglement within a quantum system may play a key role in the emergence of macroscopic irreversibility \cite{QuantumEntangle1, QuantumEntangle2, QuantumEntangle3}.

In the textbook \cite{Therm}, the second law of thermodynamics is derived from an equation known as the Fermi master equation. However, the equation itself is noted to lack a proper derivation. A modern justification for this idea is provided by the Lindblad master equation \cite{Lindblad1}, which describes systems weakly coupled to their environments. In particular, the Fermi master equation is regarded as a Born-Markov approximation of the Lindblad equation \cite{OpenSys0}. These equations have been shown to exhibit irreversible behavior, even when the underlying microscopic dynamics are unitary \cite{UniTherm1}. This approach has been applied to various models in quantum thermodynamics \cite{OpenSys1, OpenSys2, QuantChaos1, QuantChaos2}. A particularly important result of this line of research is the identification of conditions under which systems reach a steady state resembling the  single-particle distribution functions (Maxwell-Boltzmann, Fermi-Dirac or Bose-Einstein distributions) \cite{ThermalEquilibrium, QuantumThermalization1, QuantumThermalization2}.

Analogously, in Ref. \cite{2Law}, the authors derive the second law of thermodynamics from a quantum-mechanical analysis of a total system partitioned into an object and its environment. This type of theory has attracted considerable attention in recent years \cite{thermal1, thermal2, thermal3}. Additionally, the theory of non-Markovian quantum evolution \cite{NonMark1, NonMark2, NonMark3, NonMark4} has garnered significant interest. It focuses on cases where the system’s dynamics are influenced by interactions with the external world.

However, all of the aforementioned theories locate irreversibility within a subsystem of a larger, isolated system -- the environment. The total system, being isolated, is still governed by reversible equations, which represent a significant conceptual gap \cite{DecohProblem2}. In the classical Gibbs treatment of subsystems, they are considered part of a larger system in equilibrium, described by the microcanonical ensemble. Reversible equations, however, cannot reach this stable state \cite{LdStat}. From the perspective of the second law of thermodynamics, satisfying the condition of entropy growth for the entire system requires postulating a low-entropy initial state. But this is the solution, which the authors of the cited works dismiss as ad hoc and unsatisfactory.

To address this, in the present paper we use the concept of non-coherence but approach the notion from a different perspective than decoherence theory or quantum thermodynamics. Rather than focusing on the loss of coherence in a quantum system due to interaction with its environment, our investigation centers on the environment itself. We treat it as inherently non-coherent. We observe that it is the coherent states of matter that are exotic, while non-coherent states occur naturally. Coherent states -- such as superconductive and superfluid phases -- arise only in specific substances under low temperatures. Coherent light is generated by lasers. In contrast, natural light is typically non-coherent. From this, we infer that non-coherence represents the inherent, natural state of matter.

In Ref. \cite{Interface}, we proposed the concept of coherent and non-coherent transformation to describe the emergence of irreversibility associated with the transmission of quantum particles through an interface. It was suggested that if particles incident on the interface from opposite sides are phase-correlated (coherent regime), their transmission occurs reversibly. In contrast, transmission is irreversible when there is no phase correlation between particles on the two sides of the interface (non-coherent regime), which is the naturally occurring situation. An optical experiment was conducted and successfully confirmed the predictions of this theory \cite{Larotonda}. Extending this theory, we introduce the concept of coherence time into the evolution of an isolated quantum system. 

The most significant difference between the approach presented here and decoherence theory is that the latter, in the spirit of the original Copenhagen interpretation, divides changes occurring in a quantum system into two categories: normal evolution, described by the Schrödinger equation, and measurement. The latter is then associated with the process of decoherence. Of these two parts, it is measurement that is considered irreversible \cite{DecohTime}. According to this theory, decoherence arises due to the interaction of the quantum system with its environment.  We treat decoherence differently, drawing on an analogy with wave optics. In optics, non-coherence arises because no wave with a continuous spectrum is entirely monochromatic. As a result, light emitted from different sources cannot form stable interference patterns, and waves emitted from the same source will desynchronize if split and travel along paths whose length difference exceeds the coherence length \cite{Optics}. In our view, it is not the process of coherence loss that is irreversible, but rather the evolution of a system large enough that different parts of it are no longer mutually coherent.

The quantum thermodynamics approach \cite{OpenSys1} is the closest to the one we adopt here. However, we are not satisfied with this theory either. Decoherence as a result of interaction with an environment is a natural perspective in quantum technologies, where one seeks to create an entangled system of qubits and outside interactions destroy it. It is, however, a very unnatural perspective for the foundations of statistical physics, where one must artificially introduce a distinction between the system and the environment, since no such distinction is inherent to the problem at hand. Therefore, we aim to develop a theory without such a distinction.

Moreover, in Gibbs theory, the microcanonical ensemble, that is, the probability distribution of states for the closed system, serves as the starting point. It precedes the introduction of more specific distributions and even the definition of temperature. Furthermore, the second law of thermodynamics states that the entropy of an \textit{isolated system} does not decrease. Therefore, our goal is to formulate an irreversible theory of the evolution of a closed system.

To accomplish this goal, we formulate a model of an isolated quantum system described by quantum theory but with a finite spectral width of quantum states. We aim to comprehend how a system with a finite spectral width, and hence lacking coherence, evolves. The mathematical procedure used to derive equations for non-coherent light involves averaging the arguments of the complex amplitudes of waves. Such a parallel between quantum entanglement and the coherence of light was, for example, recently observed in Ref. \cite{Entanglement}. Our objective is to extend the procedure of averaging the arguments of complex amplitudes to general quantum systems and derive the equation governing non-coherent evolution.

We employ a method in this regard that is very similar to the one used in decoherence theory to eliminate the non-diagonal components of the density matrix \cite{DecohBook}. However, in our approach, the non-diagonal terms that vanish do not belong to the matrix characterizing the state of the system, but rather to the matrix describing the transformation of the system.

After performing the averaging over arguments, we obtain the equation of non-coherent evolution. We find that it coincides with the previously mentioned Fermi master equation. The Fermi master equation is equivalent to Fermi's golden rule combined with the conservation of total probability \cite{Therm}. This means that by introducing the concept of non-coherent evolution into quantum mechanics, we arrive at equations that already exist. Thus, we do not contradict any well-known or well-tested results of quantum mechanics. Rather, we re-derive and reinterpret them in a way that makes quantum mechanics compatible with statistical mechanics, particularly with the irreversibility found in kinetic equations. In \cite{Therm}, it is shown that entropy growth follows from the Fermi master equation. We will additionally show that both the equidistribution of probabilities and the Boltzmann equation in important particular cases can be readily derived from it.

\section{Non-coherent transformations}

As stated in the introduction, the concept of non-coherent transformation was introduced in Ref. \cite{Interface} to describe the transmission of quantum particles through an interface. It was proposed to explain the emergence of irreversibility in such cases. As a physical example of the formalism presented below, one can imagine a beam of quantum particles incident on an interface and being scattered in different directions. However, we will mostly not refer to this specific example and will proceed in an abstract manner.

The goal here is to develop and generalize the formalism introduced in Ref. \cite{Interface} so that it can be applied in the next section. In this section, we will first deduce the main formula for a series of non-coherent transformations and explain the origin of the argument averaging procedure. The following section will address non-coherent evolution, which is a continuous parametric family of non-coherent transformations where the parameter is time.

Let us consider a system that can occupy a set of different microstates indexed by $k$, taking values from $1$ to $N$. The system may be in different microstates with different probabilities. The probability amplitude of the system being in microstate $k$ is denoted by $A_k$, and the probability is given by $P_k = |A_k|^2$. The state of the system is characterized by the complete set of probability amplitudes. We will group all these amplitudes $A_k$ into a column.

We introduce an index-free notation for further use, as it is more transparent in certain cases. We denote matrices and column vectors using a specific font: columns of amplitudes as $\mathcal{A}$, columns of probabilities as $\mathcal{P}$, and the relation between probabilities and amplitudes is written as $\mathcal{P} = \mathcal{A}^* \circ \mathcal{A}$, where the circle denotes the Hadamard product, or element-wise product. This can be written more concisely by introducing the Hadamard square, or element-wise square:
\begin{equation}
\mathcal P = |\mathcal A|^{\circ 2}.
\end{equation}
The same notation will be applied to transformation matrices later.

Let the system evolve freely, without interactions between microstates. This implies conservation of probabilities—and hence the absolute values of amplitudes—but allows for changes in the phases of the amplitudes corresponding to different microstates. Let us denote by $\Phi$ the matrix that characterizes the phase changes. By definition, it is diagonal:
\begin{equation}
\Phi_{k j} = e^{i\varphi_k} \delta_{k j}.
\label{PhiMatrix}
\end{equation}

At some moment, the system that was initially in state $i$ undergoes changes, leading to a new state $f$, characterized by a column of amplitudes $\mathcal{A}_f$. The linear transformation between states should be unitary to conserve total probability, and we denote the unitary transformation matrix as $\mathcal{U}$. Thus, the new state of the system $\mathcal{A}_f$ is given by
\begin{align}
&\mathcal A_f = \mathcal U \Phi \mathcal A_i  \nonumber \\
&\mathcal P_f = | \mathcal U \Phi \mathcal A_i|^{\circ 2}.
\label{Gen1}
\end{align}
Let us rewrite the last formula component-wise. We denote $k_i$ indexes of microstates of the initial state and $k_f$ of the final state
\begin{equation}
P_{k_f} = |A_{k_f}|^2 = \sum_{k_i} |U_{k_f k_i}|^2 |A_{k_i}|^2 + \sum_{k'_i \neq k_i} U^*_{k_f k'_i} U_{k_f k_i} e^{i(\varphi_{k'_i}-\varphi_{k_i})} A^*_{k'_i} A_{k_i}
\label{UnitComps}
\end{equation}
We impose the non-coherence condition on the transformation, signifying that we lack knowledge of the arguments $\varphi$ of the amplitudes in Eq. \eqref{UnitComps}; only their absolute values are known. To ascertain how the probabilities transform, we average over all possible values of the arguments. To distinguish the probability calculated with non-coherence conditions imposed from the one calculated "normally" by formula \eqref{UnitComps}, we use a tilde:
\begin{equation}
\widetilde{ P_{k_f}} = \frac{1}{(2\pi)^N} \idotsint \limits_0^{2\pi} \prod_{k_i} d\varphi_{k_i} \left[ \sum_{k_i} |U_{k_f k_i}|^2 |A_{k_i}|^2 + \sum_{k'_i \neq k_i} U^*_{k_f k'_i} U_{k_f k_i} e^{i(\varphi_{k'_i}-\varphi_{k_i})} A^*_{k'_i} A_{k_i} \right].
\label{noncoherent}
\end{equation}
 We see that the integral of the sum over $k'_i \neq k_i$ equals zero since we have $\int d\varphi_{k_i} \exp i(\varphi_{k'_i}-\varphi_{k_i}) =0$. The integral over the first term equals $(2\pi)^N \sum_{k_i} |U_{k_f k_i}|^2 |A_{k_i}|^2$, since integrated function does not depend on integration variable. We see that after averaging over arguments, the column of probabilities is 
\begin{align}
& \widetilde{ P_{k_f}} = \sum_{k_i} |U_{k_f k_i}|^2 |A_{k_i}|^2   \nonumber \\
& \widetilde{\mathcal P_f} =  |\mathcal U|^{\circ 2} |\mathcal A_i|^{\circ 2}.
\end{align} 
When there are no complex-valued variables on the right-hand side of the equation, we can omit the tilde on the left, since in this case we assume non-coherence without any possible confusion. We introduce the matrix $\mathcal{T} = |\mathcal{U}|^{\circ 2}$, the matrix of non-coherent transformation and rewrite the formula as:
\begin{align}
\mathcal P_f =\mathcal T \mathcal P_i.  
\label{T1}
\end{align}
What are the physical consequences of having the transformation law \eqref{T1} instead of \eqref{Gen1}?

\begin{figure}
\includegraphics[width=0.80\textwidth]{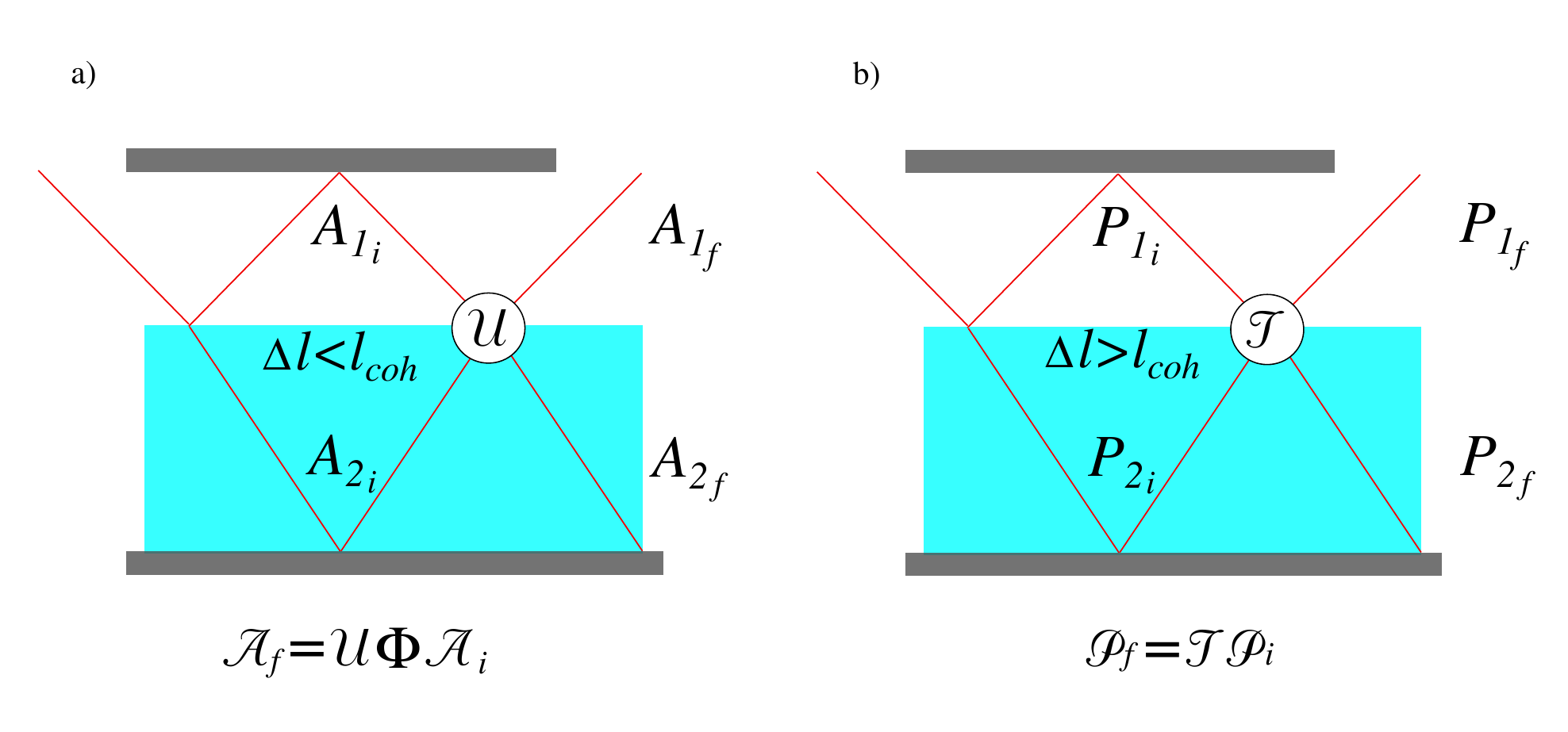}
\caption{Two beams of light are incident on the interface and are transformed either coherently or non-coherently, depending on the conditions. The interface can be, for example, between air and optical glass (shown in light blue). The gray rectangles represent mirrors. In subfigures (a) and (b), two beams are emitted by the same source and separated by the interface. The initial state of the system is defined by the amplitudes after separation. The column of initial amplitudes is denoted as $\mathcal{A}_i$, and the column of initial intensities as $\mathcal{P}_i$.\\
If the difference between the optical paths of the two split beams is smaller than the coherence length of the light, $\Delta l < l_{\text{coh}}$, a coherent transformation occurs according to formula \eqref{Gen1}, as depicted in subfigure (a). This transformation is reversible, and with appropriate phase differences between the beams, they can recombine into a single beam, meaning the splitting transformation has been reverted. If the difference between the optical paths exceeds the coherence length of the light, $\Delta l > l_{\text{coh}}$, a non-coherent transformation occurs according to formula \eqref{T1}. The resulting intensities no longer depend on the phase difference and are strictly greater than zero, which means the split beams cannot merge back into a single beam.}
\label{non-coherence-fig1}
\end{figure}

First of all, we observe that since we average over the arguments, the phases of the initial amplitudes $A_{k_i}$ play no role in the expression for the probabilities of the final state \eqref{T1}. Therefore, we do not need to consider the initial column of amplitudes $\mathcal{A}$ and can work directly with the column of probabilities $\mathcal{P}$.

Second, it follows from the properties of unitary matrices that the matrix $\mathcal{T}$ is bistochastic, meaning its elements are non-negative real numbers, and the sum of the elements in each row and each column equals one. Since the sum of the elements in each column is one, the values $T_{k_f k_i}$ can be interpreted as the probabilities of transition from state $k_i$ to state $k_f$.

The third, and most important, consequence is the irreversibility of non-coherent transformation. Bistochastic matrices generally lack inverses within the same class of matrices (except for permutation matrices, which consist solely of ones and zeroes). The physical consequence of this is that transformations described by such matrices—other than permutations—are irreversible. A detailed discussion of this is given in Ref. \cite{Interface}.

So, if we are allowed to average over arguments, we obtain irreversible behavior of the system under consideration. But what gives us the right to perform such averaging? To answer this question, we need to consider concrete physical examples.

In Ref. \cite{Larotonda}, an optical experiment is conducted to demonstrate the irreversibility of non-coherent transmission of light through the interface. First, a beam of light is split into two parts by the interface between air and optical glass. Then, after reflecting off mirrors, the two parts meet again at the interface (Fig. \ref{non-coherence-fig1}a, b). If the difference in the optical paths lengths exceeds the coherence length, $\Delta l > l_{\text{coh}}$, the resulting intensities of the beams can be calculated using a product of bistochastic matrices, without considering the phase difference between the beams. In this case, formula \eqref{T1} should be used. However, if the difference in the optical path is smaller than the coherence length, $\Delta l < l_{\text{coh}}$, the actual phase difference between the two parts of the light must be taken into account, so we calculate using formula \eqref{Gen1}. Importantly, in the latter case, if the phase difference between the two beams differs by $2\pi n$, the beams reunite into one at the interface, demonstrating the reversibility of the process.

Although the experiment in \cite{Larotonda} was carried out in the context of classical optics, the results can be interpreted in terms of photons and wave packets. First of all, the intensity of light $I$ is proportional to the number of photons $N$ multiplied by the energy flux associated with each. The flux associated with each photon is its velocity, which is the speed of light, multiplied by its energy, given by Planck's formula: $c \hbar \omega$. When photons are reflected in a given direction, the expected number is given by the product of the number of incident photons and the probability of transmission, $N P$. By the law of large numbers, the actual number of photons does not differ significantly from the expected value. Thus, we can write $I = P N c \hbar \omega$. This allows us to use intensity and probability interchangeably. We note that $P$ and $N$ are dimensionless quantities, while intensity and energy flux are synonymous and have the same dimension, so the dimensions of the left-hand side and the right-hand side of the formula are the same.

\begin{figure}
\includegraphics[width=0.90\textwidth]{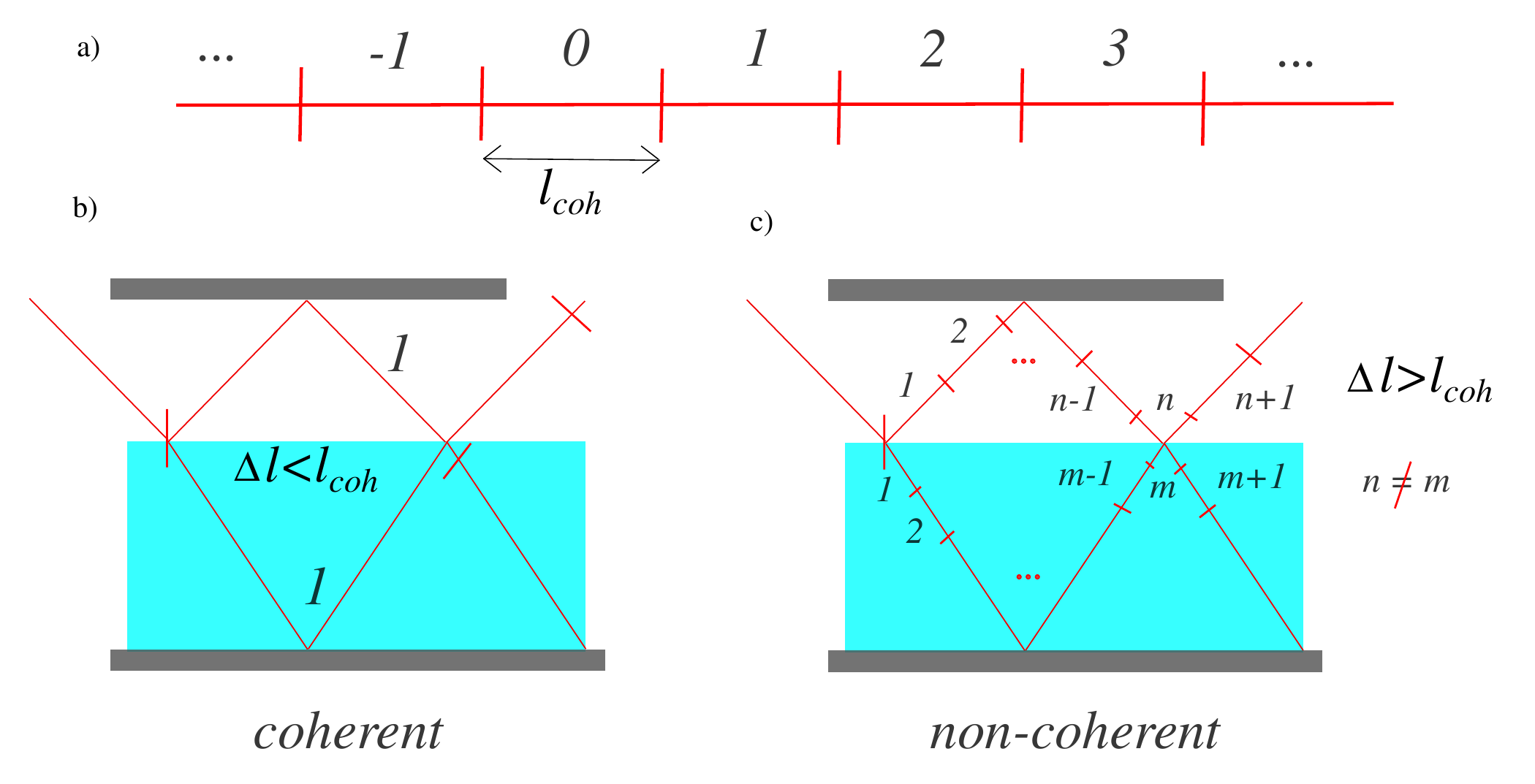}
\caption{As in Fig. \eqref{non-coherence-fig1}, two beams of light are incident on the interface and are transformed either coherently or non-coherently, depending on the conditions. We present a qualitative scheme for interpreting the experiment in terms of photon packets. Subfigure (a) shows a beam of light divided into segments of equal length—the coherence length $l_{\text{coh}}$. Each segment roughly corresponds to a single packet of photons, and the segments are numbered sequentially from left to right.\\
In subfigure (b), the left end of packet number one is shown crossing the mirror. Since the difference between the optical paths of the two split parts of the beam is smaller than the coherence length, $\Delta l < l_{\text{coh}}$, the state interacts with itself, and the interaction is coherent. In particular, it is packet number one interacting with itself. In subfigure (c), the difference between the optical paths exceeds the coherence length, $\Delta l > l_{\text{coh}}$, and the state interacts with another state, resulting in non-coherent behavior. Specifically, it is state number $n$ interacting with state number $m$, with $n \neq m$.}
\label{non-coherence-fig2}
\end{figure}

Second, the coherence length corresponds to the length of the photon wave packet, and both are inversely related to its spectral dispersion. Speaking very roughly, we can subdivide the beam into packets of photons of length $l_{\text{coh}}$, and observe that if the difference in optical paths is smaller than the size of a packet, the packet will primarily interact with itself at the interface (Fig. \ref{non-coherence-fig2}a). This type of interaction results in interference and allows a description in terms of probability amplitudes $\mathcal A$, which are transformed by the unitary matrix $\mathcal U$. Otherwise, if the size of the packet is smaller than the difference in optical paths, the state interacts with another state (Fig. \ref{non-coherence-fig2}b), and such interaction does not produce interference. In this case, photons are described by probabilities $\mathcal P$, and the transformation of these probabilities is given by the bistochastic matrix $\mathcal T$.

Let us abstract away the details of the described experiment and formulate the mathematical essence. The formulas we provide in the following few paragraphs are well-known in the contexts of classical \cite{Optics} and quantum optics \cite{Knight}, so we will not provide extensive detail. However, we aim to rewrite them in the specific context and notation of this manuscript for further use.

Also, it will be more convenient to refer to the coherence time $t_{\text{coh}}$ rather than the coherence length $l_{\text{coh}}$. The two quantities are related by the simple formula $l_{\text{coh}} = v t_{\text{coh}}$, where $v$ is the phase velocity of the wave.

If two monochromatic waves with amplitudes $A_1$ and $A_2$ overlap each other, the resulting intensity is
\begin{equation}
P_{\text{res}} = |A_1|^2 + |A_2|^2 + 2|A_1||A_2|\cos \phi, \label{fullcoh}
 \end{equation}
where $\phi$ is the difference in phase between the two waves. The term with the cosine is the interference term, which we denote as $\text{Int}$. However, we know that in practice, no wave is completely monochromatic. This means that, in frequency space, it is not a delta function, but rather a collection of many frequencies, even if they are concentrated near some given value. The amplitude at a given time is then given by the expression
\begin{equation}
A(t) = \int_{-\infty}^{\infty} d\omega e^{i\omega  t}  A(\omega).
\label{Multichrome}
\end{equation}
Considering this, the interference term is given by
\begin{equation}
\text{Int} = \int_{-\infty}^{\infty} d\omega  A_1^*(\omega)^* e^{-i\omega  t} A_2(\omega) e^{i\omega  t} = \int_{-\infty}^{\infty} d\omega  A_1(\omega)^* A_2(\omega).
\label{interference1}
\end{equation}
This formula is valid if we calculate the average intensity over a sufficiently large time interval. On smaller time scales, the intensity may fluctuate above or below this value, since the formula relies on the orthogonality of exponential functions with different frequencies -- something that is strictly true only for integration over an infinite time. However, if the averaging time interval is much greater than the inverse of the frequency difference, the integral becomes close to zero.

In the experiment described above and depicted in Fig. (\ref{non-coherence-fig1}a, \ref{non-coherence-fig1}b), the wave interacts with itself, but taken at a different moment in time. We substitute into the formula \eqref{interference1} the expressions $A_1(t) = A(t)$ and $A_2(t) = A(t+\Delta t)$, and for these we use the expression \eqref{Multichrome}, so we obtain
\begin{equation}
\text{Int}  = \int_{-\infty}^{\infty} d\omega  A^*(\omega)^* e^{-i\omega  t} A(\omega) e^{i\omega  (t+ \Delta t)} = \int_{-\infty}^{\infty} d\omega  |A(\omega)|^2  e^{i\omega \Delta t}.
\label{interference2}
\end{equation}
It can be shown that such an integral vanishes when $\Delta t \to \infty$. To gain a better understanding of the behavior of the interference as a function of $\Delta t$, we need to make some additional assumptions. The standard approach is to assume $|A(\omega)| = \mathrm{const}$ within some frequency interval, and zero outside of this interval \cite{Optics}. We denote the spectral width of the interval as $\omega_{\text{width}}$, and we can approximate the integral \eqref{interference2} as being of order one when $\Delta t < 1/\omega_{\text{width}}$, and of order $1/(\Delta t \omega_{\text{width}})$ otherwise. We define the coherence time $t_{\text{coh}}$ as $1/\omega_{\text{width}}$, and rewrite these estimations as
\begin{align}
\left[ \begin{aligned} &\Delta t < t_{\text{coh}}, \, \text{Int}  = \, 1  \\
&\Delta t > t_{\text{coh}}, \,  \text{Int}  = \frac{ t_{\text{coh}}}{\Delta t}  \end{aligned} \right.
\label{interference3}
\end{align}

We conclude that when not fully monochromatic and, generally, not fully correlated waves overlap, the resulting intensity is given by
\begin{equation} 
P_{res} = |A_1|^2 + |A_2|^2 + 2\, \text{Int} \cdot |A_1||A_2|\cos \phi,
\label{intermediate}
\end{equation} 
where $\text{Int}$ is given by expression \eqref{interference3}. We see that if $\Delta t \gg t_{\text{coh}}$, the resulting intensity is simply the sum of the intensities of the two sources. This case is referred to as the non-coherent regime.

We can explain the non-coherent regime differently. The interaction of each single-frequency component of $A_1$ in the formula \eqref{interference1} with the same-frequency component of $A_2$ is described by expression \eqref{fullcoh}. However, the phase difference $\phi$ in that formula is a function of frequency. When calculating $\text{Int}$, we integrate $\cos \phi$ over all frequencies and obtain zero.

Now that we have outlined the scheme of coherence damping between separated beams of light \eqref{interference3}, let us return to Figs. (\ref{non-coherence-fig1}, \ref{non-coherence-fig2}) and consider how different this picture is from heat bath theories, in which decoherence arises from interaction with the environment. In the scheme described here, coherence disappears without any interaction with a medium. Non-coherence results solely from the difference in the travel distances between the two parts of the beam. Nothing happens to the beam as it propagates through a nonabsorbing medium. Moreover, if we let the two separated parts of the light travel an arbitrarily long but equal optical distance before recombining them, they will remain mutually coherent.

Moreover, in a real experiment conducted in Ref. \cite{Larotonda}, there was an interaction that can be considered as an interaction between the investigated system with the thermal environment. It is an interaction between light and mirrors, and a sufficient portion of light was absorbed by mirrors (gray rectangles in Fig. (\ref{non-coherence-fig1}).  Due to losses resulting from imperfect reflection at the mirrors, we do not obtain the same intensity at the output as at the input. Therefore, the whole process is not completely reversible. The experiment demonstrated the reversibility of transmission and reflection of coherent light beams at the air–glass interface, and the non-reversibility of the same process for non-coherent light beams. In both cases, the interaction with the mirrors was equally strong, meaning that the same portion of light was absorbed. This interaction, however, did not affect the outcome of the experiment: coherent light beams remained coherent and, under appropriate phase conditions, were successfully recombined into a single beam.

Finally, let us generalize our results to an arbitrarily long series of transformations. Let the final set of probabilities be expressed as
\begin{equation}
\mathcal P_f = | \mathcal U_n \Phi_n ... \mathcal U_1 \Phi_1 \mathcal A_i|^{\circ 2},
\label{GenN}
\end{equation}
after $n$ transformations. The case where the time between initial ($i$) and final ($f$) states is much less than the coherence time of the system, means we can disregard the phase shift between transformations and thus we have
\begin{equation}
\mathcal P_f = | \mathcal U_n ... \mathcal U_1 \mathcal A_i|^{\circ 2}..
\label{UN}
\end{equation}
For the large time intervals between transformations, when the conditions for averaging over arguments are fulfilled, we can write
\begin{equation}
\widetilde{\mathcal P_f} =   \int d^N \varphi_n\, ... \int d^N \varphi_1\, | \mathcal U_n \Phi_n  ... \mathcal U_1 \Phi_1 \mathcal A_i|^{\circ 2}.
\label{GenN}
\end{equation}
where we denoted
\begin{equation}
 \int d^N \varphi_k = \frac{1}{(2\pi)^N} \idotsint \limits_0^{2\pi} \prod_{k_{j}=1}^N d\varphi_{k_{j}}
\label{IntD}
\end{equation}
 To simplify the expression \eqref{GenN} we want to write it in an index form. For \eqref{GenN} we have
\begin{equation}
P_{k_{f}} = \sum_{k_{n}, k'_{n}=1}^N ... \sum_{k_{1}, k'_{1}=1}^N \sum_{k_{i}, k'_{i}=1}^N ( U_{k_{f}, k'_n}  e^{i\varphi_{k'_{n}}} ...  U_{k'_{1}, k'_{i}} e^{-i\varphi_{i'}}  A_{k'_i})^*  ( U_{k_{f}, k_n}  e^{i\varphi_{k_{n}}} ...  U_{k_{1}, k_{i}} e^{i\varphi_{i}}  A_{k_i}).
\label{GenNComps}
\end{equation}
We perform summation over each set of microstate indeces $k_j$ for every state $j$ from $1$ to $n$.

When we do averaging over phases, couples of exponents $\exp i(\varphi_{k}-\varphi_{k'})$ turns to delta symbols $\delta_{k',k}$, so after averaging we get
\begin{equation}
\widetilde{P_{k_{f}}} = \sum_{k_{n}, k'_{n}=1}^N ... \sum_{k_{1}, k'_{1}=1}^N \sum_{k_{i}, k'_{i}=1}^N ( U_{k_{f}, k'_n} ...  U_{k'_{1}, k'_{i}} A_{k'_i})^* \delta_{k'_{n}, k_{n}} ... \delta_{k'_{i}, k_{i}}
( U_{k_{f}, k_n} ...  U_{k_{1}, k_{i}}  A_{k_i}).
\label{NDeltas}
\end{equation}
We sum over each $ k'_j$ using delta-symbols and rearrange terms
\begin{equation}
\widetilde{P_{k_{f}}} =  \sum_{k_{n}=1}^N ... \sum_{k_{1}=1}^N  \sum_{k_{i}=1}^N  U_{k_{f}, k_n}^* U_{k_{f}, k_n}   ...  U_{k_{1},k_{i}}^* U_{k_{1}, k_{i}}  A_{k_{i}}^*A_{k_{i}}.
\label{TNComps}
\end{equation}
Which, with indexless notation just gives
\begin{align}
&\widetilde{\mathcal P_f} =  |\mathcal U_n|^{\circ 2} ... |\mathcal U_1|^{\circ 2} |\mathcal A_i|^{\circ 2}  \nonumber \\
& \mathcal P_f =\mathcal T_n ... \mathcal T_1 \mathcal P_i.  
\label{TN}
\end{align}
If we compare the derived expression with \eqref{UN}, we find that \eqref{TN} involves taking Hadamard squares before multiplying the matrices, whereas \eqref{UN} involves multiplying the matrices first and then taking the Hadamard square.

Now, intuitively, if each $U_k$ represents an infinitesimal transformation, or a transformation close to the identity matrix, and there is an infinite number of such transformations, the expression \eqref{UN} would represent unitary, or coherent evolution, while the expression \eqref{TN} would represent non-coherent, or stochastic evolution.

In the next section, we will make this idea precise and combine it with basic concepts from the theory of quantum transitions under the influence of perturbations to derive the equation of non-coherent evolution.

Before proceeding to the next section, we want to highlight a serious issue in our reasoning. The Hadamard product is basis-dependent, so we must specify the basis before averaging over phases, as in \eqref{T1}. Physically, this means we must select a specific set of microstates as "true" and distinguish them from any other state that is a superposition of the chosen microstates. In this manuscript, we consider weakly interacting systems, where the choice is evident: the states of the system can be expressed by the occupation numbers of the one-particle Hamiltonian's eigenstates. As we will see later, this choice naturally leads us to the correct form of the Boltzmann equations. However, how to choose the set of "true" microstates in the general case remains unclear. This issue is already recognized in the context of decoherence theory as the superselection problem. Possible approaches to resolving it are discussed in the book \cite{DecohBook}.

\section{Non-coherent evolution}

As the previous section shows, in the discrete case, the non-coherence condition, $\Delta t \gg t_{\text{coh}}$, converts a reversible unitary transformation into an irreversible stochastic transformation. In the continuous case, unitary evolution is described by the exponential of a skew-Hermitian matrix \cite{NaiveLie}. In the context of quantum theory, we typically express this as $\mathcal{U}(t) = \exp(i \mathcal{H} t)$, where $\mathcal{H}$ is a time-independent Hermitian matrix, particularly representing the Hamiltonian of the system. The continuous analog of a bistochastic transformation is a symmetric continuous-time Markov chain (CTMC). The time dependence of transition probabilities is given by $\mathcal{T}(t) = \exp(\mathcal{Q} t)$, where $\mathcal{Q}$ is the transition rate matrix (Q-matrix) \cite{MarkovBook}. Its non-diagonal elements are positive numbers representing the rates of transition between states. The diagonal elements $Q_{\alpha\alpha}$ represent the total departure rate from the state $\alpha$. They are negative, such that the sums of elements in each column are zero\footnote{In probability texts, the state of the system is often represented by a row, and the transformation matrix is placed on the right of it. We use the convention adopted in physical texts: to represent the state as a column and place the transformation matrix to the left of it. This convention results in a difference in definition: in probability textbooks, the sums of elements in each row of a Q-matrix are zero.}. A general Q-matrix serves as the continuous analog of a stochastic transformation, while the analog of a bistochastic transformation is a Q-matrix symmetric over the main diagonal, i.e., $Q_{\alpha\beta} = Q_{\beta\alpha}$. We are going to demonstrate that just as the unitary transformation converts into a bistochastic transformation under the condition of non-coherence, unitary evolution converts into symmetric stochastic evolution.

\begin{figure}
\includegraphics[width=0.5\textwidth]{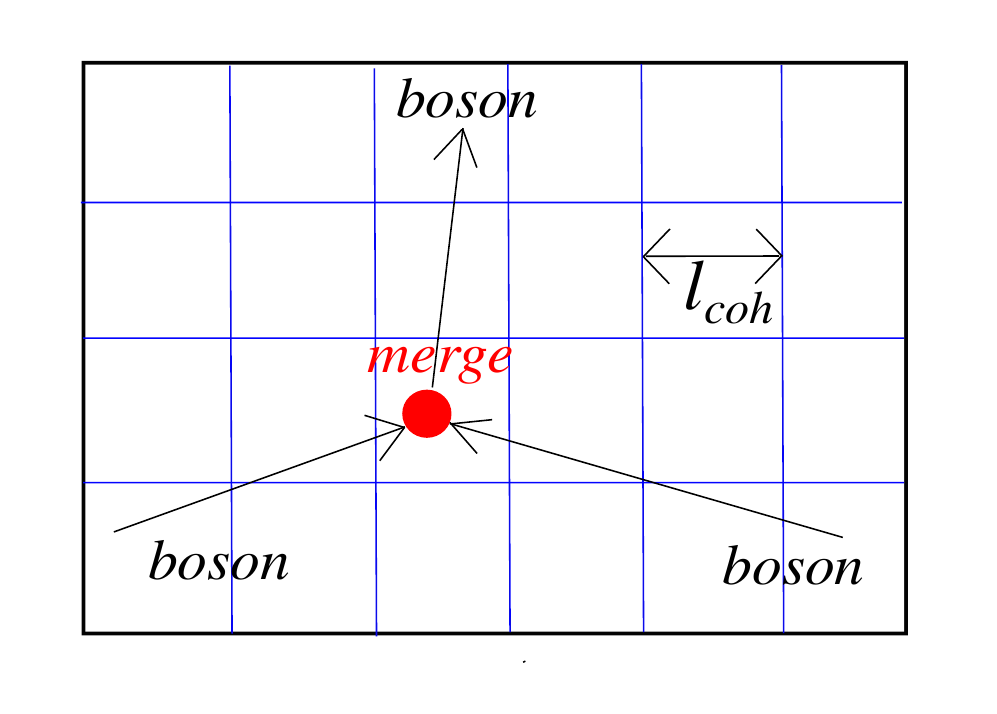}
\caption{The diagram illustrates the characteristic scales of the system. The coherence length $l_{\text{coh}}$ is much smaller than the size of the system. For example, we depict the interaction of two bosonic particles resulting in their merging into a single bosonic particle. The regions of the system from which the two particles originate are separated by a distance much greater than the coherence length, and their interaction is therefore non-coherent.}
\label{box}
\end{figure}

Let us describe a physical system to which our method will be applied. First, we consider a many-particle system described by a $\psi$-function, whose evolution is governed by the Schrödinger equation. We assume the system to be isolated and localized within a finite region of space. For the system to be of interest, its Hamiltonian must include interactions between the constituent particles.

Suppose the system has fixed energy $E$. In that case, the time dependence of its $\psi$-function is given by the exponent $e^{iEt/\hbar}$, and since the amplitudes of the wave function remain constant, the probability distribution is constant as well. However, in our treatment we want to model a non-trivial evolution of probabilities. We assume that, just as the light beam in the previous section did not consist of a single frequency but rather a collection of frequencies within some interval $\omega_{\text{idth}}$, the system under consideration does not have an exact energy $E$, but rather a collection of energies within an interval $\hbar \omega_{\text{idth}}$ around $E$. Since there is a spectral width, there is also a coherence time, defined as $t_{\text{coh}} = 1/\omega_{\text{idth}}$ (exactly as in the previous paragraph). Finally, if there is a coherence time, there is also a coherence length, $l_{\text{coh}} = v t_{\text{coh}}$, although these lengths differ for different velocities.

Now we want our system to be sufficiently large, macroscopic, and here is what we mean by this. First, a closed quantum system has a discrete spectrum, and we will make substantial use of this fact in our treatment. However, we assume that the spectral width is much greater than the typical interval between energy levels of the system. Next, we require the size of the system to be much greater than the coherence lengths corresponding to all velocities of interest.

Ideally, we would like to prove that the evolution of such a system already possesses the desired properties: namely, that it is irreversible and that its limiting state corresponds to an equidistribution of probabilities over all microstates of the system. However, the methods we have in mind do not permit a solution in full generality. Therefore, we introduce two additional simplifying assumptions. First, we assume that the interaction between particles is weak. Second, we assume that the system is homogeneous. Both of these assumptions will be clarified further in the text.

Let us now give a qualitative description of why this system should be irreversible. To this end, we compare it with the light beam investigated in the previous section. There, we subdivided the beam into imaginary segments of length $l_{\text{coh}}$ and found that when photons from different segments are incident on the interface simultaneously, the transformation of associated energy flux is irreversible (Fig. \ref{non-coherence-fig2}). In the case of the light beam, all photons move in the same direction with the same speed, so in order for photons from different regions to meet, we had to split the beam into two parts with paths of different optical lengths.

In analogy, let us make an imaginative subdivision of the quantum system described earlier into zones of characteristic size $l_{\text{coh}}$. Unlike the beam, the waves that constitute our system propagate in all directions and with different velocities. Consequently, waves originating from different zones naturally cross and interact. However, for our argument to hold, the interaction must be weak enough that a typical interaction involves zones separated by distances much larger than the coherence length (Fig. \ref{box}). This is precisely the role of the weak-interaction assumption.

With this qualitative picture in mind, let us now present the mathematical formalism that corresponds to the intuitive idea. We start similarly to a classic simple time-dependent perturbation theory \cite{LdQuant} that is aimed to derive Fermi's golden rule. We seek for a wave function of a many-body system in the form
\begin{equation}
|\Psi \rangle = \sum_k A_k |k\rangle,
\label{Psi}
\end{equation}
where $|k\rangle$ are states of non-interacting system, and $A_k$ are probability amplitudes of system to be in state $|k\rangle$. We get the following equation for  $A_k$:
\begin{equation}
i \hbar \sum_k \dot A_k |k\rangle = \sum_l \exp\left(\frac{i}{\hbar}(E_k - E_l)t \right) \hat V  A_l |l\rangle.
\label{smth}
\end{equation}
We introduce $V_{kl} = \langle k|\hat V|l\rangle$ and get the equation for amplitudes
\begin{equation}
 \dot A_k  = -\frac{i}{\hbar} \sum_l V_{kl} \exp\left(\frac{i}{\hbar}(E_k - E_l)t \right) A_l,
\label{Start}
\end{equation}

We want to make two important observations on properties of Eq. \eqref{Start}. First, this equation describes unitary evolution, since on a time interval that tends to zero, we denote it as $dt$, the change of the column of amplitudes is given by
\begin{equation}
\mathcal{A}(t+dt) = e^{\mathcal{M}(t) dt} \mathcal{A}(t), 
\label{TimeDep}
\end{equation}
where elements of matrix $\mathcal{M}(t)$ are defined by expression in the right side of \eqref{Start}:
\begin{equation}
M_{kl}(t) =  -\frac{i}{\hbar} V_{kl} \exp\left(\frac{i}{\hbar}(E_k - E_l)t \right).
\label{TimeDep2}
\end{equation}
We see that evolution described by Eq. \eqref{TimeDep} is unitary since $M_{kl}$ is skew-Hermitian: $M_{kl} = - M_{lk}^\dag$. On a finite time interval, transformation of amplitudes can be expressed as an infinite product of exponents of the form \eqref{TimeDep}, so it is unitary as a product of unitary transforms.

Second, if a diagonal element of the interaction matrix is nonzero, $V_{kk} \neq 0$, which means self-interaction, we can make a substitution $A'_k = A_k \exp(\frac{i}{\hbar} V_{kk} t)$, so that the diagonal term in Eq. \eqref{Start} cancels out. Introducing exponential multipliers in such a form to the wave function \eqref{Psi} is equivalent to changing the energy of state $|k\rangle$ by $V_{kk}$. This coincides with the correction to energy in the first order of time-independent perturbation theory. Further, we will assume energies of states already include self-interaction and $V_{kk} = 0$.

Equation \eqref{Start} is too complicated to tackle directly. To continue our investigation, we need to use an approximate method. The original Dirac's approach is to employ the variation of parameters method. We choose a different approach here.

Let us assume that the amplitudes of the wave function are slowly changing over time compared to the fast oscillating exponent $\exp\left( i(E_k - E_l)t/\hbar \right)$ in \eqref{Start}, for most pairs of values of $k$ and $l$. We want to distinguish the slowly oscillating terms of $\mathcal{M}(t)$ with $E_k \approx E_l$ that contribute to the change of amplitudes, from the fast oscillating terms that should affect only the fast oscillating part of the wave function. To do that, we integrate over some technical timescale $\Delta t$ that is small compared to the typical timescale of change of $A_k$, which we denote as $T_A$. However, $(E_k - E_l)\Delta t/\hbar \gg 1$ for most values of $k$ and $l$.

\begin{figure}
\includegraphics[width=0.99\textwidth]{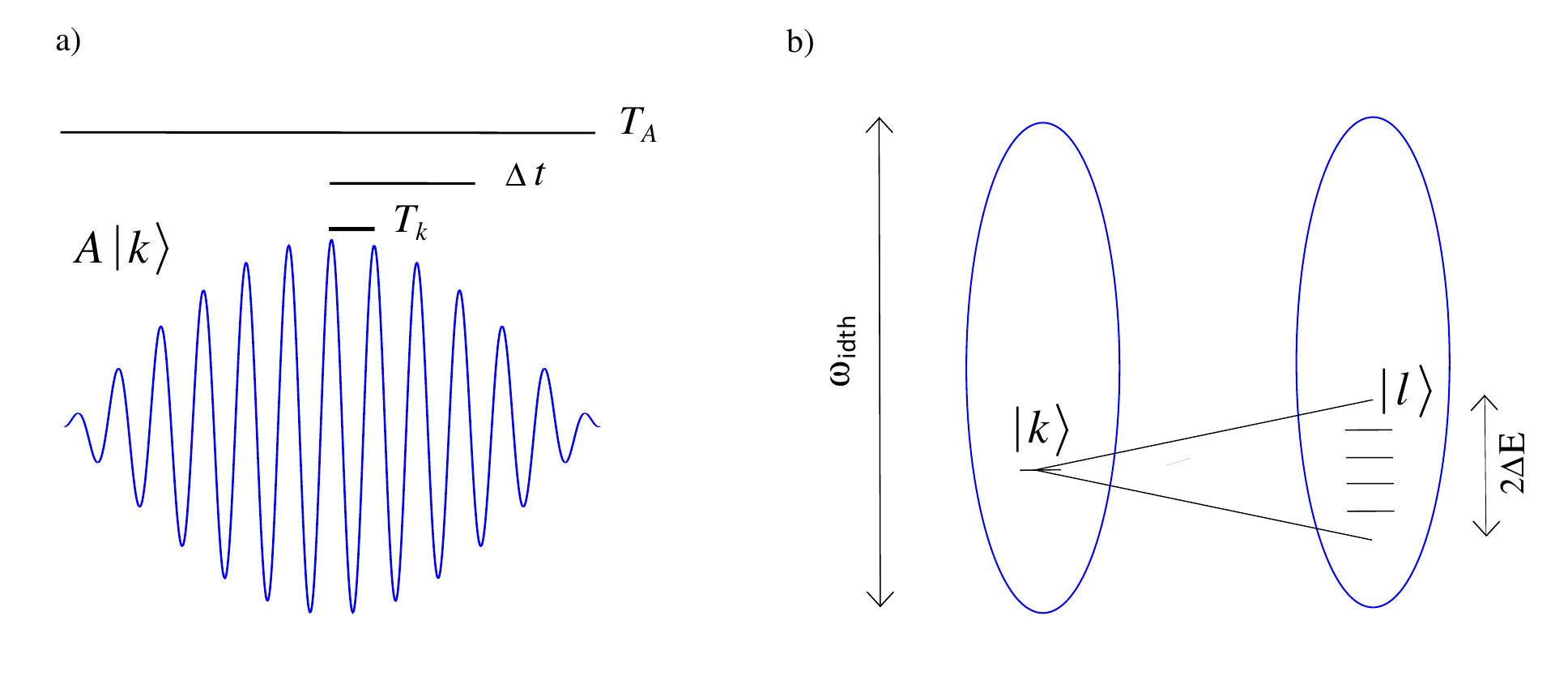}
\caption{The diagram illustrates the characteristic scales of transitions in a non-coherent regime. To make the diagram easier to understand, quantities that are an order of magnitude greater than others are shown as only slightly greater. Both the time scale and energy scale are depicted, as some proportions are more intuitive on the timescale, while others are more intuitive on the energy scale. The quantities on the time scale and energy scale are related by:\\
$T_k = \frac{\hbar}{E_k}$, $t_{\text{coh}} = \frac{1}{\omega_{\text{idth}}}$, $\Delta t = \frac{\pi\hbar}{\Delta E}$, $T_A = \frac{\hbar}{V_M}$.\\
\textbf{(a)} shows a microstate $|k\rangle$ and the probability amplitude of the system to be in such a microstate, $A$. It demonstrates that the characteristic scale of the amplitude evolution, $T_A$, is much higher than the working scale, $\Delta t$, which is much higher than the wave period of a state, $T_k$. Therefore, we have:\\
$T_k \ll \Delta t \ll T_A$.\\
\textbf{(b)} depicts a sheaf of states (represented as an oval), with a frequency width $\omega_{\text{idth}}$, containing a single state $|k\rangle$ in the first sheaf and a group of states in the second sheaf, indexed by $l$, such that the energy difference between them and the initial state is less than $\Delta E$. The width of the sheaf is much greater than the working broadening, $\Delta E$. It is evident that the number of states to which a transition can occur from the initial state $|k\rangle$ is proportional to the broadening $\Delta E$.}
\label{scales}
\end{figure}

Our general strategy for the subsequent derivation is to use the technical timescale $\Delta t$ as an intermediate value between the characteristic timescales of the fast-changing part of the wavefunction $|k\rangle$ and the slowly varying envelope $A_k$ (Fig. \ref{scales}a). When integrating the microstate wavefunction $|k\rangle$ over $\Delta t$, we treat $\Delta t$ as a large quantity, almost infinity. However, when we deal with the amplitude $A_k$, we treat $\Delta t$ as a small quantity, nearly zero. Nevertheless, in both cases, we cannot simply take the limits $\Delta t \to 0$ or $\Delta t \to \infty$, because we would lose important information by doing so. We need to keep it finite and take the limits only at the very last steps of our derivation. In the end, we will see that if such an intermediate timescale exists, the results we obtain do not depend on the specific value of $\Delta t$.

After integration over  $\Delta t$, on the left side of the equation \eqref{Start}, we obtain $A(\Delta t) - A(0)$. On the right side,
\begin{align}
\int \limits_0^{\Delta t} \exp\left(\frac{i}{\hbar}(E_k - E_l)t \right) A_l\, dt \approx 
A_l \int \limits_0^{\Delta t} \exp\left(\frac{i}{\hbar}(E_k - E_l)t \right) dt = 
- \frac{i\hbar}{E_k - E_l}\left(e^{\frac{i}{\hbar}(E_k - E_l) \Delta t} - 1 \right) A_l.
\label{smoothing}
\end{align}

We divide both sides by $\Delta t$. We want to take a closer look at the obtained dimensionless expression before $A$ on the right-hand side of the equation. We define
\begin{align}
\chi_{kl} = - \frac{i\hbar}{(E_k - E_l) \Delta t}\left(e^{\frac{i}{\hbar}(E_k - E_l) \Delta t} - 1 \right).
\end{align}
This expression has two limiting cases:
\begin{align}
|E_k - E_l|  \gg \hbar/\Delta t =>\chi_{kl}(E_k) \approx 0, \nonumber \\
|E_k - E_l|  \ll \hbar/\Delta t => \chi_{kl}(E_k) \approx 1.
\label{LimsChi}
\end{align}

We see that $\chi_{kl}$ guarantees that only states with energies close to $E_k$ contribute to the variation of $A_k$. Physically, it is connected with the time–energy uncertainty principle: we cannot distinguish energies that differ by less than $\Delta E$ over a time shorter than $\hbar/\Delta E$.

 We want to emphasize this point even more and simplify our calculation. To do that, we replace $\chi_{kl}$ with an indicator function $\overline\chi_{kl}$, which equals one in a region of energies near $E_k$ and zero outside of this region. We call the energy radius $\Delta E$: 
\begin{align}
|E_k - E_l| > \Delta E & => \overline \chi_{kl} = 0 \nonumber \\
|E_k - E_l| < \Delta E & => \overline \chi_{kl} = 1.
\label{DefOverChi}
\end{align}
We want to choose the value of $\Delta E$ so that
\begin{equation}
\int dE |\chi_{kl}|^2 = \int dE |\overline\chi_{kl}|^2.
\label{Condition}
\end{equation}
Introducing integration variable $E = (E_k - E_l)$
\begin{equation}
\int dE |\chi_{kl}|^2 = \frac{4 \hbar^2}{\Delta t^2} \int dE \frac{\sin^2(E\Delta t/2\hbar)}{E^2} = 2\pi\hbar/\Delta t.
\label{Chi1}
\end{equation}
From the definition of the indicator function
\begin{equation}
\int dE |\overline\chi_{kl}|^2 = \int dE |\overline\chi| = \int_{-\Delta E}^{\Delta E} dE = 2 \Delta E.
\label{Chi2}
\end{equation}
Substituting Eqs. (\ref{Chi1}, \ref{Chi2}) to \eqref{Condition} we find that $\Delta E = \pi\hbar/\Delta t$. Comparing this with Eq. \eqref{DefOverChi} we get
\begin{align}
|E_k - E_l| > \pi\hbar/\Delta t & => \overline \chi_{kl} = 0 \nonumber \\
|E_k - E_l| < \pi\hbar/\Delta t & => \overline \chi_{kl} = 1.
\label{LimsOverChi}
\end{align}
Condition \eqref{Condition} is hard to justify at this point. At the end of the derivation, we will see why it is necessary; see the discussion after Eq. \eqref{Fermi}. There, we explain why the exchange of $\chi_{kl}$ with $\overline\chi_{kl}$ is possible and what information we have lost in doing so.

 On the left side of the equation \eqref{Start}, after performing transformations, we have 
\begin{align}
\frac{A_k(\Delta t) - A_k(0)}{\Delta t} \approx \dot A_k,
\label{derivative}
\end{align}
but here we have a time derivative of smoothed $A$, which we have truncated the fast oscillating terms from.

We come to the equation for amplitudes, that are smooth, slowly oscillating part of the total wave function:
\begin{equation}
 \dot A_k  = -\frac{i}{\hbar} \sum_l \overline \chi_{kl} V_{kl} A_l.
\label{Main}
\end{equation}
Later we will not write $\overline \chi_{kl}$ explicitely, but will keep in mind that we sum over $l$-s such that $E_k - E_l < \Delta E$ (Fig. \ref{scales}b). In the matrix form the equation is
\begin{equation}
\dot{\mathcal A} =  -\frac{i}{\hbar} \mathcal V  \mathcal A
\label{Matrix2}
\end{equation}
This equation is much simpler than the one we started from in \eqref{Start}, because the matrix elements are now time-independent. However, we note that the matrix $\mathcal{V}$ is Hermitian, as we can see from the explicit form \eqref{Main}, so equation \eqref{Matrix2} still describes unitary evolution of the system. Thus, we have not lost this fundamental property through our simplifications.

Using equation \eqref{Matrix2}, we can estimate the characteristic timescale of the evolution of amplitudes, which we have denoted as $T_A$. We take the smallest nonzero element of the matrix $\mathcal{V}$ and denote it as $V_M$. Then the minimum possible rate of change of an amplitude is proportional to $V_M$, and we can write $T_A = \hbar / V_M$ (Fig. \ref{scales}a).

The solution to Eq. \eqref{Matrix2} is 
\begin{equation}
\mathcal A(t) =  \exp \left(-\frac{i}{\hbar} \mathcal V t \right) \mathcal A(0)
\label{Solution}
\end{equation}

We may present the exponent as an infinite product and write
 \begin{equation}
\mathcal A(t) = \lim_{n \to \infty} \left( \mathcal E -\frac{i}{\hbar} \mathcal V t/n \right)^n \mathcal A(0).
\label{ExpDef}
\end{equation}
Here $\mathcal E$ denotes the identity matrix.

The strategy of our derivation is to take $n$ as a large number, but not instantly take the limit $n \to \infty$. If we take this limit we would get $\dot P = 0$. Instead, we want to expand the exponent in Eq. \eqref{Solution} in a Taylor series, keeping terms up to the first non-vanishing order in $1/n$, so we obtain a small change over the time $t/n$:
\begin{equation}
\mathcal A(\Delta t) =  \left(\mathcal E -\frac{i}{\hbar} \mathcal V \Delta t - \frac{1}{2 \hbar^2} \mathcal V^2 \Delta t^2 + O(\Delta t^3)  \right) \mathcal A(0),
\label{SmallCange}
\end{equation}
 Where we denote $\Delta t = t/n$. Soon we will see the importace of keeping expression of order $\Delta t^2$.

Then we multiply $n$ small changes to obtain the evolution over an arbitrary time interval $t$:
 \begin{equation}
\mathcal A(t) =  \left(\mathcal E -\frac{i}{\hbar} \mathcal V \Delta t - \frac{1}{2 \hbar^2} \mathcal V^2 \Delta t^2 + O(\Delta t^3)  \right)^n \mathcal A(0).
\label{ExpProd}
\end{equation}
 In the end we will analyze the asymptotic behavior of obtained expression at $n \to \infty$. In this way, we have rewritten formula \eqref{Solution}.

We now want to analyze the non-coherent regime. This means that the state under consideration, $|k\rangle$, belongs to a sheaf of states with a width in the frequency region $\omega_{\text{idth}}$, and we choose a working timescale, $\Delta t$, such that it is much larger than the coherence time. Equivalently, we have $\Delta E \ll \hbar \omega_{\text{idth}}$ (Fig. \ref{scales}b). The quantity $\Delta E$ fits into $\hbar \omega_{\text{idth}}$ many times, each time with a different phase difference for each pair of amplitudes, so we can average over the phase. This allows the formula for a series of non-coherent transformations, given in \eqref{TN}, to be applicable. We note that this approximation is in close analogy with the Born-Markov approximation of the Lindblad equation \cite{OpenSys0}, where it is assumed that the heat bath relaxation time is small compared to the relaxation time of the system.

Let us write a product formula for $\mathcal P$
 \begin{equation}
\mathcal P(t) =  \mathcal T^n \mathcal P(0),
\label{ExpProdT}
\end{equation}

 Similarly to what we have done with unitary matrices in expression \eqref{ExpDef}, we want to express bistochastic matrices $\mathcal T$ in a form of unit matrix plus $\mathcal Q \Delta t$. Using \eqref{T1} we have
 \begin{equation}
 \mathcal T = \mathcal E + \mathcal Q \Delta t = |\mathcal E -\frac{i}{\hbar} \mathcal V \Delta t |^{\circ 2},
\label{SmallT}
\end{equation}

For the non-diagonal terms of $\mathcal Q$ we find that
\begin{equation}
k \neq l, \ \ \ Q_{kl} = \frac{1}{\hbar^2} |V_{kl}|^2 \Delta t .
\label{QNonDiag}
\end{equation}
For diagonal terms the identity matrices in Eq. \eqref{SmallT} cancel out, and we have to calculate the diagonal of the matrix
\begin{equation}
Q_{kk} = \left( \frac{\Delta t}{\hbar^2} \right) \left(\frac{1}{2 } \mathcal V^2 + \frac{1}{2} \mathcal V^{*^2}\right)_{kk},
\label{QPreDiag}
\end{equation}
which is obtained by multiplying the second-order terms in $\Delta t$ from Eq. \eqref{SmallCange} with the identity matrices. For this we use the fact that $\mathcal V$ is Hermitian with zero diagonal elements, so we obtain 
\begin{equation}
Q_{kk} = - \frac{1}{\hbar^2} \left( \sum_l |V_{kl}|^2 \right) \Delta t.
\label{QDiag}
\end{equation}
We can see that this satisfies the conditions for the transition rate matrix of a CTMC. The non-diagonal elements are positive numbers, and the diagonal elements are negative, such that the sums of the elements in each column are zero. We observe that if we had not retained elements of order $\Delta t^2$ in \eqref{ExpProd}, we would have zeros on the main diagonal of $\mathcal{Q}$.

We rewrite the formula \eqref{ExpProdT} as $\exp(\mathcal{Q} t)(1 + O(1/n))$. Now, recalling that $\Delta t = t/n$ is a small quantity compared to the characteristic timescale of change of $\mathcal{P}$, we have $1/n \ll 1$, and we can discard it. Thus, we obtain:
\begin{equation}
\mathcal P(t)  = \exp(\mathcal Q t)  \mathcal P(0).
\label{StochExp}
\end{equation}
Which means $P(t)$ is the solution of the differential equation
\begin{equation}
\dot{\mathcal P} =  \mathcal Q  \mathcal P.
\label{StochEvol}
\end{equation}
This is the equation of non-coherent evolution that we were looking for.

We can reason differently. Instead of using the solution for $\mathcal{A}(t)$ to construct $\mathcal{P}(t)$ and then deducing the equation for $\mathcal{P}(t)$, we aim to derive the differential equation directly. Since the probability of the system being in state $k$ is $P_k = A_k^* A_k$, its time derivative is normally given by $\dot{P}_k = \dot{A}_k^* A_k + A_k^* \dot{A}_k$. However, we use the smoothed derivative of the amplitude, as given by \eqref{derivative}, which is not defined at a single point but uses two different points in time. This introduces an ambiguity: we can substitute $A_k(0)$ or $A_k(\Delta t)$ into the formula, or their linear combination, or even take the value of the amplitude at any point within the interval between 0 and $\Delta t$, and this yields different answers for $\dot{P}_k$. To overcome this difficulty, we use the smoothed derivative of the probability, just as we used the smoothed derivative of the amplitude in \eqref{derivative}. We regroup the terms so that the formula resembles the normal one based on the product rule, while also being symmetric with respect to the transposition of $A_k^*$ and $A_k$. We omit the $k$ index and write
\begin{align}
\dot P =  \frac{P(\Delta t) - P(0)}{\Delta t} = \frac{A^*(\Delta t)A(\Delta t) - A^*(0) A(0)}{\Delta t} = \nonumber \\
= \frac{[A^*(\Delta t)-A^*(0)]A(\Delta t)/2 + A^*(0)A(\Delta t)/2 + A^*(\Delta t)[A(\Delta t) - A(0)]/2 +A^*(\Delta t)A(0)/2 - A^*(0) A(0)}{\Delta t} = \nonumber \\
= \dot A^* \frac{A(\Delta t) + A(0)}{2} +  \frac{A^*(\Delta t) + A^*(0)}{2}  \dot A.
\label{ProbDif}
\end{align}
We note that the first and the last expressions are equivalent, both can be used for calculation. We introduce the last form because it resembles the product rule formula. And we are going to use the last form in subsequent manipulations.

For $\mathcal{A}(\Delta t)$ we use the formula \eqref{SmallCange}. This time we want to write up the computation element-wise, to present it both ways. For $A_k(\Delta t)$, only keeping terms of the first order in $\Delta t$:
\begin{equation}
A_k(\Delta t) = A_k(0) - \frac{i}{\hbar} \sum_l V_{kl} \Delta t A_l(0)
\label{Delta2}
\end{equation}
Then for derivative, to the first non-vanishing term in $\Delta t$ 
\begin{equation}
\dot A_k(\Delta t) =  - \frac{i}{\hbar} \sum_l V_{kl} A_l(0) + \frac{1}{\hbar^2} \left(\sum_l |V_{kl}|^2 \right) \Delta t A_k(0).
\label{DeltaD}
\end{equation}
We can write equivalent expressions for $A^*_k$ and its derivative. 

We substitute expressions (\ref{Delta2}, \ref{DeltaD}) and their complex conjugates into \eqref{ProbDif} and imply non-coherence condition that makes $A^*_k A_l \rightarrow \delta_{kl} |A_k|^2 = P_k$. Thus we obtain 
\begin{equation}
\dot P_k = - \frac{1}{\hbar^2} \left( \sum_l |V_{kl}|^2 \right) \Delta t P_k + \frac{1}{\hbar^2} \sum_l |V_{kl}|^2 \Delta t P_l.
\label{TransitionRates}
\end{equation}
Taking into consideration the definition of the $\mathcal{Q}$ matrix (\ref{QNonDiag}, \ref{QDiag}), we see that it is the same equation as \eqref{StochEvol}. We observe that we obtain the same result with a slightly different reasoning. In the latter case, the connections with the results of the previous section are less apparent, but the differential equation arises more naturally.

Let us now address the issue with formula \eqref{StochEvol} (or its equivalent \eqref{TransitionRates}): its dependence on the arbitrarily chosen parameter $\Delta t$. We will demonstrate that, with the chosen approximations, the system's dynamics do not depend on the choice of $\Delta t$.

We begin by utilizing an approximation of a continuous spectrum. We can do it since, as we have assumed at the beginning of the section, the spectral width of the system is much greater than the energy difference between the levels of the system. Expression $E(l) = \text{const}$, defines an energy surface in the space of quantum numbers $l$. Near a given state $l$, we divide the quantum numbers into two components. The first component, $l_T$, is tangential to the energy surface; in other words, small changes in its value leave the energy unchanged. The second component, $l_N$, is normal to the energy surface. Additionally, we recall expression \eqref{LimsOverChi} and substitute $\Delta t = \pi \hbar/\Delta E$ into the second term on the right-hand side of Eq. \eqref{TransitionRates}. We obtain:
\begin{equation}
\frac{1}{\hbar^2} \sum_l |V_{kl}|^2 \Delta t P_l = 
\frac{\pi}{\hbar \Delta E} \sum_{l_T, l_N } |V_{kl}|^2  P_{l_T, l_N} 
\label{ToDE}
\end{equation}
We suppose $V_{kl}$ and $P_{l}$ are smooth functions of energy at the point of interest, so in a short interval of values of $l_N$ we can approximate them as constants. So summation over $l_N$ gives the number of states inside the energy region near $l$ -- $N(l)$ (Fig. \ref{scales}b):
\begin{equation}
\frac{\pi}{\hbar \Delta E} \sum_{l_T, l_N } |V_{kl}|^2  P_{l_T, l_N } = 
\frac{\pi N(l)}{\hbar \Delta E} \sum_{l_T} |V_{k\, l_T}|^2  P_{l_T} =
\frac{2 \pi}{\hbar} \sum_{l_T} |V_{k\, l_T}|^2  P_{l_T} \nu(l)
\label{Fermi0}
\end{equation}
where $\nu(l) = N(l)/(2\Delta E)$ is the density of states near $l$. Multiplier 2 arises because $\Delta E$ is the absolute value of the deviation of energy from $E_l$ and the energy in the region can be less than $E_l$ and higher than $E_l$ so the length of the region is  $2\Delta E$.

We can reason differently. We perform the same subsitution $\Delta t = \pi \hbar/\Delta E$ as in \eqref{ToDE}, and approximate the sum as an integral (we recall that we have omitted $\overline \chi_{kl}$ in the expression for the summation for brevity):
\begin{equation}
\frac{1}{\hbar^2} \sum_l |V_{kl}|^2 \Delta t P_l = \frac{2 \pi}{\hbar} C \int dl \frac{\overline \chi_{kl}}{\Delta E} |V_{kl}|^2 P_l,
\label{IntForm}
\end{equation}
where $C$ is some constant with dimension $[1/l]$ that arises due to the transition from the sum to the integral. We want to illustrate this transition from summation to integration with an example commonly used in solid-state physics \cite{LdKin}. It is the transition from summation over momenta $p$ to integration over them. We use a one-dimensional example for simplicity: $\sum_p \to \frac{L}{2 \pi \hbar} \int dp$. Here $L$ is the length of the system under consideration. Typically, it cancels out in the final result. In this example, the constant $C = \frac{L}{2 \pi \hbar}$.

 We observe that $(\overline \chi_{kl})/(2 \Delta E)$ is a function that equals $1/(2 \Delta E)$ in the energy region near $E_K$ with width $\Delta E$, and is zero elsewhere. As $\Delta E \to 0$, the region of energies shrinks to zero, but the value of the transition rate grows to infinity. We can see that the integral of $(\overline \chi_{\Delta E})/(2 \Delta E)$ is a function that equals $1/(2 \Delta E)$ over energy, and its total integral is one. Thus, in the weak limit, the function approaches $\delta(E - E_k)$. Therefore, for transmissions between neighborhoods of states $k$ and $l$, we obtain:
\begin{equation}
Q_{kl} = \frac{2 \pi}{\hbar} |V_{kl}|^2 \delta(E_k - E_l).
\label{Fermi}
\end{equation}
That finishes our derivation.

At this point we can understand what we have gained by changing $\chi_{kl}$ to $\overline \chi_{kl}$ (see Eqs. \ref{LimsChi}, \ref{LimsOverChi}). If not for this exchange, we would have to keep the function $\chi_{kl}$ inside the summations over $l$ after Eq. \eqref{Main}. We would have to carry it through all the manipulations concerning non-coherence. The function $\chi_{kl}$ is "glued" to the interaction $V_{kl}$, so when we get the squares of the latter, it would be combined with the former: $|\chi_{kl}V_{kl}|^2$. So in place of expression \eqref{IntForm} we would have:
\begin{equation}
\frac{1}{\hbar^2} \sum_l |\chi_{kl}V_{kl}|^2 \Delta t P_l = \frac{2 \pi}{\hbar} C \int dl \frac{|\chi_{kl}|^2}{\Delta E} |V_{kl}|^2 P_l.
\label{IntForm}
\end{equation}
Because of condition \eqref{Condition}, the subsequent argument holds and we obtain exactly the same result \eqref{Fermi}. The argument that appeals more to visual intuition and leads to the form of equation \eqref{Fermi0} does not hold with the much more complicated function $\chi_{kl}$.

While the exchange of $\chi_{kl}$ for $\overline \chi_{kl}$ allows us to use shorter formulas in the computation and permits more intuitive reasoning, there is very important information that we lose with this exchange. Before we take the square of $\chi_{kl}$, it behaves qualitatively differently from $\overline \chi_{kl}$. The limit of $(\chi_{\Delta E})/(2 \Delta E)$ at $\Delta E \to 0$ is not just a delta function, but a sum $\delta(E - E_k)+ \text{v.p.} \frac{i}{E - E_k}$, where $\text{v.p.}$ means that taking an integral over $E$ one should take its principal value. This expression is very well-known in the context of perturbation theory. The second term is responsible for perturbational corrections to the values of energies of states and matrix elements of interaction. When we interchange $\chi_{kl}$ for $\overline \chi_{kl}$, we discard any perturbative effects of interactions.

We could have chosen a more rigorous method for deriving equation \eqref{Matrix2}. We would still seek a smoothed unitary evolution equation, which we achieve by expressing the matrix $\mathcal{V}$ in the form of the Dyson series. If we only retain the first term of the series for $\mathcal{V}$, it gives the same expression that we have found. Seemingly, this method allows us to find higher-order terms of the perturbation series, including corrections to energy levels due to interactions and second-order transitions. However, this approach is lengthy, and if we were to take this route, we would aim to investigate perturbation theory within the described framework.In this paper, our primary goal is to deduce the non-coherent evolution equation from the smoothed unitary evolution equation \eqref{Matrix2}. The interchange of $\chi_{kl}$ for $\overline \chi_{kl}$ is a shortcut to achieve this goal. We leave the task of finding the most accurate form of the smoothed unitary evolution equation, which includes perturbative corrections, for subsequent publications.

Our analysis began with partitioning the wave function of the system into two components: fast-varying time-dependent wavefunctions of microstates and slow-varying probability amplitudes for the system to be in a given state, as shown in \eqref{Psi} (Fig. \ref{scales}a). It is important to note that the distinction between coherent and non-coherent evolution regimes can only be made at the level of smooth amplitudes. The microstates we consider are combinations, either symmetric or anti-symmetric, of the exact stationary solutions to the single-particle Schrödinger equation. At this level, we cannot even introduce the concepts of coherence or non-coherence. The situation changes when we consider envelopes that contain a sheaf of states within a small energy interval and the interactions between them (Fig. \ref{scales}b). As Eq. \eqref{interference3} shows, sheaves can be phase-correlated, which gives the regime of unitary evolution \eqref{Matrix2}, or completely uncorrelated, which leads to the regime of non-coherent evolution \eqref{StochEvol}.

We see that if we do not consider the full set of equations \eqref{StochEvol} describing the evolution of the system's state through transitions between microstates, but instead focus on the single transition rate \eqref{Fermi}, we recover the well-known Fermi rule. Thus, we re-derived the formula for calculating transition rates between states. 

Since the Bohr model of the atom, the notion of quantum jumps has been part of quantum theory. The term is somewhat ironic, as in the Bohr model these transitions between states were assumed to occur instantaneously. In decoherence theory, by contrast, quantum jumps are understood as a continuous process of local decoherence occurring on a very short timescale \cite{NoJump}. In our derivation, transitions appear instantaneous only when compared with the technical timescale $\Delta t$. In principle, the techniques developed here could allow one to estimate the actual transition time, though doing so would require a different set of ideas than those presented in this manuscript.

\section{Applications to statistical mechanics}

In the previous section, we deduced the equation of stochastic evolution and, by considering a single element of the transition matrix, derived the Fermi rule. We could reason in the opposite direction, as was done in Ref. \cite{Therm}. If we combine all the transition rates into a matrix and add the diagonal elements to satisfy the probability conservation condition, we obtain a stochastic equation known as the Fermi master equation. It coincides with the non-coherent evolution equation \eqref{TransitionRates}. This provides a possible way to deduce Eq. \eqref{TransitionRates} without reference to non-coherence. However, it lacks proper justification.

We want to use the derived equation to address the problem of irreversibility in statistical mechanics. We consider all three instances of this problem mentioned in the introduction: the second law of thermodynamics, the equidistribution of states in equilibrium, and the irreversibility of the Boltzmann equation. 

For the secon law we use the Gibbs entropy defined as
\begin{equation}
S = - \sum_k P_k \ln P_k 
\label{Entropy}
\end{equation}
We note that we do not use von Neumann entropy or other types of quantum entropies  \cite{OpenSys0}. The von Neumann entropy has the advantage of being invariant under transformations of basis. However, in our treatment, as we explained at the end of Section II we already work in a specified basis. Also, the goal of our manuscript is to derive the Gibbs statistical mechanics, so we use Gibbs entropy.  

In \cite{Therm}, it is proven that entropy defined by formula \eqref{Entropy} increases in a system whose evolution is described by the Fermi master equation, unless it has already reached its maximum. 
Gibbs entropy attains its maximum when the probabilities of all states are equal. Thus, this also proves the equiprobability of states in equilibrium. However, we can approach the problem from the opposite direction.

As we mentioned earlier, the equation of non-coherent evolution \eqref{StochEvol} is a continuous-time Markov chain (CTMC) equation. It is known \cite{MarkovBook} that such equations are irreversible and converge to a stationary distribution for large values of $t$. The specific distribution depends on the particular $\mathcal{Q}$-matrix. However, the $\mathcal{Q}$-matrix of non-coherent evolution, defined by expressions (\ref{QNonDiag}, \ref{QDiag}), is symmetric. For symmetric $\mathcal{Q}$-matrices, the stationary distribution is always the same: an equiprobability of all microstates.

The symmetry of the $\mathcal{Q}$-matrix aligns with the principle of detailed balance. This symmetry is a consequence of the Hermiticity of the interaction matrix $\mathcal{V}$.

In contrast with the general case proven in \cite{MarkovBook}, the symmetric case is simpler, and we can directly verify that the equiprobability of all microstates is a stationary solution of equation \eqref{StochEvol}. We substitute $P_k = 1/N$ for each $k$ into the system \eqref{StochEvol}, where $N$ is the number of equations (i.e., the number of microstates). Since the sum of the elements in each row of the symmetric $\mathcal{Q}$-matrix equals zero, the right-hand side of \eqref{StochEvol} vanishes. Therefore, the time derivatives of all $P_\alpha$ equal zero. This shows that the distribution $P_\alpha = 1/N$ is stationary, and regardless of the initial condition, in the limit $t \to \infty$, the probability of each state becomes equal. This means the distribution to which the non-coherent evolution of a closed macroscopic system converges is the Gibbs microcanonical ensemble.

We note that entropy growth is a stronger property of non-coherent evolution than irreversibility alone. All CTMCs are irreversible; however, only symmetric ones have the equiprobability of states as a stationary solution. Moreover, equiprobability corresponds to the maximum of entropy as defined by equation \eqref{Entropy}. Thus, entropy growth is a consequence of irreversibility combined with the principle of detailed balance.

We also emphasize that the approximation \eqref{LimsOverChi} neglects the perturbation of states due to interaction. We may speculate that even when interactions are taken into account, the system still undergoes stochastic evolution described by equation \eqref{StochEvol}, but in terms of microstates with renormalized energies and interaction matrix elements. This conjecture is motivated by the fact that Gibbs' theory remains valid for systems with strong interactions. Since we have shown that equation \eqref{StochEvol} leads to the microcanonical ensemble, it is plausible that for general systems the equidistribution of microstate probabilities results from the evolution equation being a symmetric CTMC. However, we are not in a position to prove this rigorously, and therefore we proceed under the assumption of weak interaction, for which approximation \eqref{LimsOverChi} remains valid.

Now we want to derive the Boltzmann collision integral for two particular cases. Let us consider the first case: the transitions between energy levels by electrons with the emission or absorption of phonons. More generally, it could be any fermions and bosons. The interaction has the form $V_{ij} \hat{c}^\dagger_i \hat{c}_j \hat{a}$, where $\hat{c}, \hat{a}$ are annihilation operators for fermions and bosons, respectively, and indices $i, j$ denote fermionic single-particle states.

Let us first consider just two fermionic states and the transitions between them. We denote states of a noncoherent system by angular brackets on both sides; in this notation, we consider transitions $\langle0, 1, n-1\rangle \leftrightarrow \langle1, 0, n \rangle$. Numbers in the brackets are the occupation numbers of fermionic states 1 and 2, respectively, and the third one is the number of bosons. We use formula \eqref{TransitionRates} with $k = \langle0, 1, n-1\rangle$, $l = \langle1, 0 ,n \rangle$, and $q = 2\pi|V_{1, 2}|^2 \nu/\hbar$:
\begin{align}
\dot P_{\langle0, 1, n-1\rangle} = -nq P_{\langle0, 1, n-1\rangle} + nq P_{\langle1, 0 ,n \rangle} \nonumber \\
\dot P_{\langle1, 0 ,n \rangle} = -nq P_{\langle1, 0 ,n \rangle} + nq P_{\langle0, 1, n-1\rangle}.
\label{BoltzStart}
\end{align}

We want to find how the mean occupation number (which is a one-particle distribution function) of the first state $f_1$ varies in time. We write down the expression $f_1$ through probabilities
\begin{align}
f_1 = &P_{\langle0, 1, 0\rangle}\times0 + P_{\langle0, 1, 1\rangle}\times0 + P_{\langle0, 1, 2\rangle}\times0 + ...  \nonumber \\
+ P_{\langle1, 0, 0\rangle}\times1 +  &P_{\langle1, 0, 1\rangle}\times1 + P_{\langle1, 0, 2\rangle}\times1 + ...
\label{mean}
\end{align}
Vertically aligned are states between which the transitions occur. 

We take time derivative from both sides and using \eqref{BoltzStart} we get
\begin{align}
\dot f_1 =  (-qP_{\langle1, 0, 1\rangle} + qP_{\langle0, 1, 0\rangle}) + (-2qP_{\langle1, 0, 2\rangle} + 2qP_{\langle0, 1, 1\rangle}) + ... =  
\sum_n nq (P_{\langle0, 1, n-1\rangle} - P_{\langle1, 0 ,n \rangle}).
\label{rate}
\end{align}

To deduce the one-particle distribution function from many-particle probabilities, we need to make a simplifying assumption. We assume that the probabilities of different occupation numbers of different states are independent of each other. That means $P_{\langle l, m, n\rangle} = P_{\langle l, ...\rangle}P_{\langle ..., m\rangle}P_{\langle n\rangle}$. For the mean occupation number of phonons $N$ we have 
\begin{align}
N = \sum_{n=0} n P_{\langle n\rangle}
\label{MeanPhon}
\end{align}
We substitute this expressions into \eqref{rate} and obtain:
\begin{align}
\dot f_1 = \sum_n nq (P_{\langle0, ...\rangle}P_{\langle ..., 1\rangle}P_{\langle n-1\rangle} - P_{\langle1, ... \rangle} P_{\langle ..., 0\rangle}P_{\langle  n\rangle})
= q \sum_n n\left[ (1-f_1) f_2 P_{\langle n-1\rangle} - f_1(1-f_2) P_{\langle  n\rangle}) \right] =  \nonumber \\
= q \left[ (1-f_1) f_2 \sum_{k=0} (k+1) P_{\langle k\rangle} - f_1(1-f_2) \sum_{k=0} k P_{\langle k\rangle}\right] = q \left[ (1-f_1) f_2 (N+1) - f_1(1-f_2)N \right].
\label{rate}
\end{align}
 Summation over all possible transitions yields the Boltzmann collision integral \cite{LdKin}. 

Let us derive the collision integral for the case of the three-phonon process. We consider three bosonic states, with occupation numbers $n_1, n_2, n_3$. The sum of the energies of the first two states is equal to the energy of the third state, and interaction permits bosons of the first two states to merge into a boson in a third state and vice versa. So transitions under consideration are
 $ \langle n_1-1, n_2-1, n_3+1 \rangle  \leftrightarrow  \langle n_1, n_2, n_3\rangle \leftrightarrow \langle n_1+1, n_2+1, n_3-1\rangle$. With formula \eqref{TransitionRates} we have:
\begin{align}
\dot P_{\langle n_1, n_2, n_3\rangle} = q n_1 n_2 (n_3+1) P_{\langle n_1-1, n_2-1, n_3+1\rangle} + q (n_1+1) (n_2+1) n_3 P_{\langle n_1+1, n_2+1, n_3-1 \rangle} - \nonumber \\
 q \left(  n_1 n_2 (n_3+1) +  (n_1+1) (n_2+1) n_3  \right) P_{\langle n_1, n_2, n_3\rangle} .
\label{3phonStart}
\end{align}
We take the time derivative of Eq. \eqref{MeanPhon} and substitute Eq. \eqref{3phonStart} into it to get
\begin{align}
\dot N_1 = q \sum_{n_1, n_2, n_3} n_1 [  n_1 n_2 (n_3+1) P_{\langle n_1-1, n_2-1, n_3+1\rangle} + (n_1+1) (n_2+1) n_3 P_{\langle n_1+1, n_2+1, n_3-1 \rangle} - \nonumber \\
 \left(  n_1 n_2 (n_3+1) +  (n_1+1) (n_2+1) n_3  \right) P_{\langle n_1, n_2, n_3\rangle}  ]
\label{rate2}
\end{align}
We make independence assumption and sum over $n_2, n_3$ using formula \eqref{MeanPhon} and shifting indices similarly to what we did with expression \eqref{rate}, to get
\begin{align}
\dot N_1 = q \sum_{n_1} n_1 [  n_1  P_{\langle n_1-1\rangle}(N_2 +1) N_3 +  (n_1+1) P_{\langle n_1+1 \rangle} N_2 (N_3+1) - \nonumber \\
  n_1  P_{\langle n_1\rangle} N_2 (N_3+1) - (n_1+1)  P_{\langle n_1\rangle}  (N_2+1) N_3  ]
\label{rate2}
\end{align}
We regroup summands, take $n_1$ into brackets, shift indices and obtain
\begin{align}
\dot N_1 = q \sum_{n_1} \left( (N_2+1) N_3 [  (n_1+1)^2  - n_1 (n_1+1)]  + 
   N_2 (N_3+1) [  (n_1-1)n_1 - n_1^2 ] \right) P_{\langle n_1\rangle}   
\label{near}
\end{align}
Summation over $n_1$  yeilds
\begin{align}
\dot N_1 =  q (N_1+1)  (N_2+1) N_3 - q N_1 N_2 (N_3+1)
\label{finish}
\end{align}
Summation over all possible combinations of the second and third states gives the first part of the collision integral for the three-phonon processes. The second part is obtained similarly if we consider the decay of phonons in state one into two phonons of states two and three and vice versa. It gives contribution of the form \\$q (N_1+1) N_2 N_3 - q N_1 (N_2+1) (N_3+1)$. The calculation is completely analogous \cite{LdKin}. 

\begin{figure}
\includegraphics[width=0.6\textwidth]{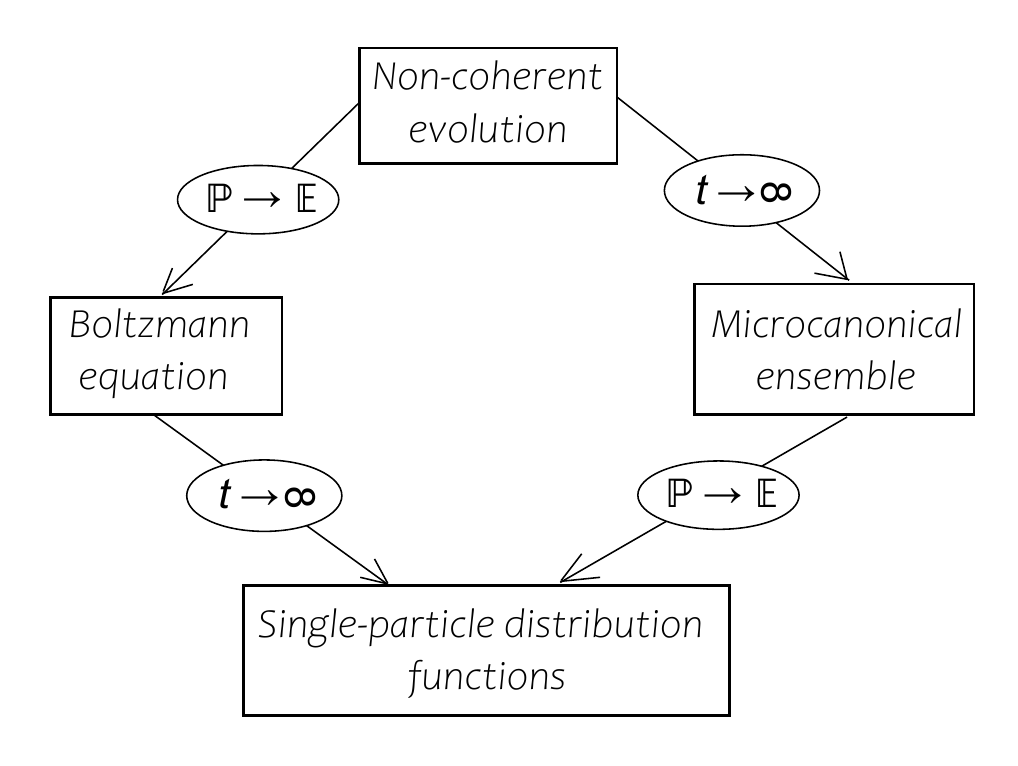}
\caption{The diagram shows how the Boltzmann equation and the Gibbs microcanonical ensemble are deduced from the equation of non-coherent evolution (Eq. \ref{TransitionRates}). In rectangles, the names of equations are indicated, and arrows represent the mathematical procedures used to derive a formula from a more general one. $t \to \infty$ denotes taking the stationary point of the equation, and $\mathbb{P} \to \mathbb{E}$ indicates averaging -- that is, the transition from probabilities to average (expected) values. The diagram is commutative in the sense that taking the stationary point of Eq. \eqref{TransitionRates} and then averaging, or averaging first and then taking the stationary point, both lead to well-known single-particle distribution functions (Fermi-Dirac or Bose-Einstein distributions). }
\label{diagram}
\end{figure}

Thus, we have derived two cases of the Boltzmann collision integral. In this derivation, the irreversibility of the Boltzmann equation is no surprise, since the irreversibility of the collision term is inherited from the irreversibility of the non-coherent evolution equation \eqref{StochEvol}, from which it is derived. This stands in sharp contrast to the BBGKY approach to the derivation of the Boltzmann equation \cite{BBGKY}. In BBGKY, irreversibility emerges as a consequence of the transition from many-particle to single-particle distribution functions, whereas in the present theory the evolution of the many-particle distribution function is already irreversible.

After performing the smoothing operation \eqref{smoothing}, only contributions from waves with frequencies much larger than $1/\Delta t$ are retained, while information about low-frequency components is entirely suppressed. We propose that such low-frequency modes contribute to the force term in the Boltzmann equation and should therefore be treated separately. In the present work, we have derived the interaction within a local region, which corresponds to the collision term in the Boltzmann equation. By contrast, the free propagation of particles, corresponding to the diffusion term, and the coupling to low-frequency fields, corresponding to the force term, must be obtained through an independent treatment.

To this end, we suggest extending the formalism by allowing the amplitudes of the psi-function \eqref{Psi} to depend slowly on the spatial coordinates, in addition to their slow temporal variation. Under this assumption, the non-collision contributions to the Boltzmann equation can be derived consistently, with low-frequency fields entering explicitly as a force term. In the present paper, however, we restrict attention to the homogeneous case, where the amplitudes are taken to be spatially uniform and external low-frequency fields are absent. This is the precise meaning of 'homogeneity' as introduced in the list of assumptions at the beginning of Section III.

In this section, we have derived the Gibbs microcanonical ensemble, which forms the basis for all equilibrium statistical mechanics, but exclusively in the case of weakly interacting systems, or simply gases. We also derived the collision part of the Boltzmann equation, which is the fundamental equation of the physical kinetics of gases. It is interesting to note that the one-particle distribution functions of quantum gases (Fermi-Dirac or Bose-Einstein distributions) can be derived in two ways: by finding the mean occupation number from the microcanonical ensemble or by finding the stationary point of the Boltzmann collision integral. In the present paper, we have derived the Gibbs microcanonical distribution as a stationary point of the equation of non-coherent evolution (Eq. \ref{TransitionRates}), and the Boltzmann collision integral as the result of the transition from the evolution equation for probabilities to the equation for average values. This means that if we start from Eq. \ref{TransitionRates}, we can first take the stationary point and then perform the averaging, or first take the averages and then find the stationary point. Both approaches lead to the same one-particle distribution functions. This result is depicted in the diagram in Fig. \ref{diagram}.

We want to add that typically, the canonical ensemble is derived from the microcanonical ensemble, and then everything in equilibrium statistical mechanics is derived from the canonical ensemble. However, the one-particle distribution functions can be derived directly from the microcanonical ensemble (see Ref. \cite{Kubo}). In the canonical ensemble, the probabilities of different occupation numbers of different states are inherently independent of each other. During the derivation from the microcanonical ensemble, this assumption is introduced, in the same way we did in our derivation of the Boltzmann collision integral.

\section{Summary and discussion}

The present work develops a theoretical framework connecting quantum mechanics with irreversible statistical processes through the concept of non-coherent evolution. We have demonstrated how averaging over phases in quantum systems leads to the emergence of irreversible behavior, providing a bridge between reversible quantum mechanics and irreversible physical kinetics. We also provided an explanation of how this averaging arises. The most important feature of this paper is the explanation of irreversible behaviour in the isolated system, which explains the second law of thermodynamics and the microcanonical ensemble as an equilibrium state of the system.

We can loosely divide the manuscript into two parts. In the first part (Section 2 and  Section 3), we derive the equation of non-coherent evolution. In the second part (Section 4), we use it to resolve problems associated with the irreversible behaviour of macroscopic systems.

At the beginning, we considered non-coherent transformations, showing how they naturally arise from unitary transformations when phase information is lost. Non-coherence arises from the fact that no wave has exactly one frequency but is composed of a sheaf of waves with closely spaced frequencies. Such sheaves can be mutually coherent or non-coherent.  These mechanism was illustrated using optical systems, where direct experimental verification is possible.

Then we move on from studying a discrete series of transformations to continuous evolution. We derive the non-coherent evolution equation, which is the central result of this work (\ref{StochEvol}, \ref{TransitionRates}). Our analysis begins with partitioning the wave function of the system into two components: fast-varying psi functions of microstates and probability amplitudes, which are smooth (in-time) envelopes of the psi function \eqref{Psi}. Our derivation relies on the assumption of weak interaction and the existence of well-separated time scales. We introduced a technical timescale $\Delta t$, chosen between the microstate oscillation period and the characteristic scale of amplitude variation. Using this technical assumption, we showed that for timescales much larger than the coherence time, unitary evolution transforms into symmetric stochastic evolution, described by continuous-time Markov chains. We then show that the final results of the calculations do not depend on the value of the timescale $\Delta t$.

Importantly, our derivation reveals the connection between non-coherent evolution and Fermi's golden rule, showing how the latter emerges naturally from our more general framework. Fermi's rule calculates constant transition rates. Constant transition rates are a property of a stochastic process. For unitary evolution, in contrast, transition rates depend on the current values of amplitudes, as can be seen from Eq. \eqref{Matrix2}. The unitarity of evolution, presented in the starting equation \eqref{Start}, is implicitly broken during the classical derivation of Fermi's rule. With our methods, we explicitly identify the juncture in our derivation where the unitarity of the equations is broken.

There is also one dubious point in the classical derivation of Fermi's rule: we integrate over an infinite time period to obtain a momentary transition rate. The introduction of the technical timescale makes this point very clear. This timescale is very large -- "infinite" -- compared to the microstate oscillation period, and very small -- "infinitesimal" -- compared to the characteristic scale of amplitude variation. We can treat it accordingly at different points in the derivation. Thus, we believe that the presented theory deepens our understanding of quantum transitions.

We continue by showing that the equation of non-coherent evolution serves as a common basis for both equilibrium statistical mechanics and kinetic theory. We derived the Gibbs microcanonical ensemble as its stationary solution, under the assumption of weak interaction, and showed that this stationary distribution corresponds to the equiprobability of microstates. We also derived the Boltzmann collision integrals in two important cases by transitioning from equations for probabilities to equations for average values. This derivation justifies the irreversibility of the Boltzmann equation as a direct consequence of the irreversibility of non-coherent evolution. We note that the Boltzmann collision integral is derived strictly from the non-coherent evolution equation and basic definitions, without any "hand-waving".

A notable property of this derivation is the commutativity of taking average values and taking the stationary point of the evolution. If we perform these operations, we arrive at the single-particle distribution functions (Fermi-Dirac or Bose-Einstein distributions), and it does not matter in which order the operations are carried out. This highlights a deeper connection between the methods of equilibrium statistical mechanics and physical kinetics.

In the introduction, we briefly outlined existing theories of irreversibility and indicated how the approach in this manuscript differs from them. Now that we have presented our findings, let us revisit this comparison in more detail.

First, consider the measurement problem and its treatment in decoherence theory. While our theory does not address the problem of outcomes, we argue that no definite outcomes are required to derive statistical mechanics from quantum mechanics. This is because only expectation values are needed, and in statistical mechanics we deal with very large systems, where the law of large numbers guarantees that actual outcomes do not differ significantly from the expected values.

 Furthermore, since the early days of quantum theory, it has been suggested that measurement results from the interaction of a quantum system with a macroscopic body.  We understand a macroscopic body as one to which the laws of classical statistical physics, as developed by Gibbs and Boltzmann, can be applied. We hope that clarifying what constitutes a macroscopic body, without appealing to interactions with the environment, will allow us to escape the vicious circle: the definition of macroscopic bodies through decoherence, and the definition of decoherence as the result of interaction with a macroscopic body. A clearer understanding of what qualifies as a macroscopic body may then shed light on what happens when a coherent quantum system interacts with it, and thus open a new perspective on the measurement problem. 

Second, let us turn to quantum thermodynamics. While it is true that the equation of non-coherent evolution (\ref{StochEvol}, \ref{TransitionRates}) is in many respects analogous to those found in quantum thermodynamics, we hope it is clear that both the physical picture offered by the present theory and the mathematical methods employed are very different. Concepts such as coherence length are not introduced in quantum thermodynamics. We may hope that further development of the theory will provide a testable and decisive distinction between the two approaches.

We have already emphasized the distinction of the present approach, which treats non-coherence as an intrinsic property of the system rather than as a consequence of interaction with the environment. Shifting the focus away from interaction is the most radical departure from almost all existing theories -- not only those based on quantum mechanics and decoherence, but even those rooted in classical mechanics, such as BBGKY hierarchy \cite{BBGKY}. All the mentioned theories in their core attribute irreversibility to interaction. Indeed, the theories based on classical mechanics, roughly speaking, attribute irreversibility to chaotization, which is a result of interatomic collisions. In our theory, in contrast, interaction is not a cause of irreversibility.

If it were not for non-coherence, interactions such as the emission of a boson by a fermion (used as an example throughout this text) would occur reversibly, as described by equation \eqref{Matrix2}. A single interaction would place the system into a superposition of states with and without the emitted boson. It is in direct analogy with our optical experiment in coherent regime, where photons after an incidence at an interface are placed into a superposition of being reflected and transmitted (Fig. \ref{non-coherence-fig2}b). The full set of all interactions would then yield a superposition of many microstates, each with a specific amplitude. Such a giant superposition evolves unitarily, according to \eqref{Matrix2}, and is therefore reversible, even when all interactions are included.

It is only when we impose the condition of non-coherence that the situation changes. The system then reduces to one in which all microstates have definite probabilities but lack definite phases, analogous to photons in the non-coherent experiment (Fig. \ref{non-coherence-fig2}c). The evolution of such a system is governed by the stochastic equation (\ref{StochEvol}, \ref{TransitionRates}), which is irreversible. Thus, in the present viewpoint, it is not interaction but non-coherence that constitutes the true origin of irreversibility.

Now, how can we test the theory? After all, in this paper we started from a well-known equation (the Schrödinger equation) and, after some derivations, arrived at well-known results (the microcanonical ensemble, the Boltzmann collision integral). The experiment presented in Ref. \cite{Larotonda} does serve as a test: it shows that light, traditionally described by reversible wave equations, exhibits irreversible properties due to non-coherence. Yet this experiment is probably too simple to be fully convincing.

One possible way of testing the theory would be, in some sense, the opposite experiment: namely, to identify wave-like behavior in transport phenomena and, crucially, to devise a method for measuring the coherence length in such systems. We suggest that transport in nanostructures offers a promising opportunity, since the characteristic scale of these systems may be smaller than or comparable to the coherence length. The essential task would be to vary system parameters in order to observe the transition from coherent to non-coherent transport, under conditions where the particle mean free path remains much larger than the structural scale. Demonstrating such a transition would establish that the mean free path and the coherence length are distinct physical quantities, thereby providing unambiguous evidence that the loss of coherence cannot be attributed to interactions.

Transport in nanostructures is a rapidly developing field \cite{NanoWires1, NanoWires2, Graphene}, and the distinction between coherent and non-coherent transmission has already been discussed in relation to interfaces \cite{NanoCoherent1, NanoCoherent2}. It is therefore quite possible that some effects consistent with the present theory are already reported in the literature. Otherwise, we may hope to design a realistic experiment to verify the theory in the near future.

From a theoretical perspective, an important investigation direction is to study systems that are neither fully coherent nor fully non-coherent, systems in which the coherence length must explicitly enter the results of calculations. This is equally relevant in the context of optics and in the context of transport theories at the nanoscale. While coherence forms a spectrum, ranging from fully coherent through partially coherent to non-coherent (see Eq. \ref{interference3}), reversibility is a binary property: a process is either reversible or irreversible. To fully understand irreversibility, one must analyze the (ir)reversible behavior of systems in which coherence is only partially lost.

Another important direction is to extend the scope of the theory, as was briefly outlined in the manuscript: namely, to incorporate perturbations of energies and transition rates due to interactions, and to generalize the framework to non-homogeneous cases.

\acknowledgments{The author would like to thank Miguel Larotonda, who patiently answered my questions about coherence in the context of classical and quantum optics, and whose insightful responses greatly improved my understanding of these areas.} 
 
During the preparation of this work, the author used ChatGPT in order to improve the quality of English. After using this service, the author reviewed and edited the content as needed and takes full responsibility for the content of the publication.

%%%%%%%%%%%%%%%%%%


\begin{thebibliography}{01}
\bibitem{LdStat} L.\,D. Landau and E.\,M. Lifshitz, \textit{Course of Theoretical Physics, Vol. 5: Statistical Physics Part 1} (Pergamon, New York, 1980) {[}Chapter 1, \S 8; Chapter 1, \S 5{]}.
\bibitem{Prig} D. Kondepudi, I. Prigogine  \textit{Modern Thermodynamics: From Heat Engines to Dissipative Structures} (Wiley, New York, 2014);  {[}Chapter 3, \S 4{]}.
\bibitem{Time1}  H.\,D. Zeh,  \textit{The Physical Basis of the Direction of Time} (Springer-Verlag, Berlin, 1999).
\bibitem{Time2} A.\,L. Kuzemsky,  \textit{The Mystery Of Time: Asymmetry Of Time And Irreversibility In The Natural Processes} (World Scientific, New Jersey, 2022).
\bibitem{Kubo}  R. Kubo, N. Hashitsume, T. Usui, H. Ichimura, \textit{Statistical Mechanics: An Advanced Course with Problems and Solutions} (Elsevier Science, 1990, Amsterdam){[}Chapter I, \S 3; Chapter I, Problem 31{]}.
\bibitem{BBGKY} S. Harris, \textit{An introduction to the theory of the Boltzmann equation} (Dover Publications, New York, 2004).
\bibitem{MeanField1} F. Golse, \textit{On the mean field and classical limits of quantum mechanics} Commun. Math. Phys. \textbf{343}, 165 (2016).
\bibitem{MeanField2} F. Golse, \textit{Mean-Field Limits in Statistical Dynamics}, Lecture notes, arXiv:2201.02005 (2022).
\bibitem{Gibbs} S. Popescu, A.\,J. Short, A. Winter Nature Phys. \textit{Entanglement and the foundations of statistical mechanics} \textbf{2}, 754 (2006).
\bibitem{Universe} R.\,M. Wald, \textit{The arrow of time and the initial conditions of the universe} Stud. Hist. and Phil. of Sci. \textbf{37},  394 (2006).
\bibitem{NofTime} S.\,M. Carroll and J. Chen, \textit{Does Inflation Provide Natural Initial Conditions for the Universe?},  Int. J. Mod. Phys. D, \textbf{14}, 2335 (2005).
\bibitem{AdHoc} E.\,G.\,D. Cohen and T.\,H. Berlin, \textit{Note on the Derivation of the Liouville Equation from the BBGKY Hierarchy}, Physica (Amsterdam) \textbf{26}, 717 (1960).
\bibitem{Modify} G.\,C. Ghirardi, A. Rimini, and T. Weber, \textit{Unified dynamics for microscopic and macroscopic systems}, Phys. Rev. D \textbf{34}, 470 (1986).
\bibitem{Therm} J.\,R. Waldram, \textit{The Theory of Thermodynamics} (Cambridge University Press, Cambridge, England, 1985).
\bibitem{Lindblad1} G. Lindblad, \textit{On the generators of quantum dynamical semigroups}, Commun. Math. Phys. \textbf{48}, 119 (1976).
\bibitem{OpenSys0} H.-P. Breuer, and F. Petruccione, \textit{The Theory of Open Quantum Systems} (Oxford University Press, 2007)  {[} Chapter 3, \S 2; Chapter 3, \S 3;  Chapter 2, \S 3{]}.
\bibitem{UniTherm1} S.\,K Pal, C.\,L. Sriram, S. Gupta, \textit{Emergent thermalization thresholds in unitary dynamics of inhomogeneously disordered quantum systems}, Phys. Rev. E \textbf{113}, 014205 (2026).
\bibitem{OpenSys1} P.\,L. Krapivsky, S. Redner, and E. Ben-Naim, \textit{A Kinetic View of Statistical Physics} (Cambridge University Press, 2013).
\bibitem{OpenSys2} A. Rivas, S.\,F. Huelga, \textit{Open Quantum Systems: An Introduction} (Springer, Berlin, 2012).
\bibitem{QuantChaos1} S. Pilatowsky-Cameo, C.\,B. Dag, W.\,W. Ho, and S. Choi,  \textit{Complete Hilbert-space ergodicity in quantum dynamics of generalized Fibonacci drives}, Phys. Rev. Lett. \textbf{131}, 250401 (2023).
\bibitem{QuantChaos2} I. Vallejo-Fabila, A.\,K. Das, D.\,A. Zarate-Herrada, A.\,S. Matsoukas-Roubeas, E.\,J. Torres-Herrera, and L.\,F. Santos, \textit{Reducing dynamical fluctuations and enforcing self-averaging by opening many-body quantum systems}, Phys. Rev. B \textbf{110}, 075138 (2024).
\bibitem{ThermalEquilibrium} M. Le Bellac, \textit{Thermal Field Theory} (Cambridge University Press, 2000).
\bibitem{QuantumThermalization1} M. Rigol, V. Dunjko, and M. Olshanii, \textit{Thermalization and its mechanism for generic isolated quantum systems}, Nature \textbf{452}, 854 (2008).
\bibitem{QuantumThermalization2} I. Vallejo-Fabila, F. Borgonovi, F.\,M Izrailev, L.\,F Santos,  \textit{Thermalization in the mixed-field Ising model: An occupation number perspective}, arXiv preprint arXiv:2601.02497 (2026).
\bibitem{Quantum} L. Maccone, \textit{Quantum solution to the arrow-of-time dilemma}, Phys. Rev. Lett. \textbf{103}, 080401 (2009).
\bibitem{DecohTime} L.\,S. Schulman, \textit{Time’s Arrow and Quantum Measurement}
(Cambridge University Press, Cambridge, England, 1997).
\bibitem{DecohBook} E. Joos, H.\,D. Zeh, C. Kiefer, D. Giulini, J. Kupsch, I.-O. Stamatescu  \textit{Decoherence and the Appearance of a Classical World in Quantum Theory} (Springer, Berlin, 2003); {[}Chapter 2, Chapter 3, Chapter 6{]}.
\bibitem{DecohProblem1} O. Pessoa, \textit{Can the Decoherence Approach Help to Solve the Measurement Problem?} Synthese \textbf{113}, 323 (1997).
\bibitem{DecohProblem2} S.\,L. Adler, \textit{Why Decoherence Has Not Solved the Measurement Problem: A Response to P.\,W. Anderson}, Stud. Hist. and Phil. of Mod. Phys \textbf{34}, 135 (2003).  
\bibitem{DecohProblem3} D. Wallace,  \textit{Decoherence and its Role in the Modern Measurement Problem}, Phil. Trans. R. Soc. A. \textbf{370}, 4576  (2012).
\bibitem{QuantumEntangle1} M. Horodecki, P. Horodecki, R. Horodecki, and K. Horodecki,  \textit{Quantum entanglement}, Rev. Mod. Phys. \textbf{81}, 865 (2009).
\bibitem{QuantumEntangle2} C.\,Y Zhang, S.\,X Zhang, Z.\,X Li, \textit{Bond Additivity and Persistent Geometric Imprints of Entanglement in Quantum Thermalization}, arXiv preprint arXiv:2601.01327 (2026).
\bibitem{QuantumEntangle3} D. Jennings and T. Rudolph, \textit{Entanglement and the Thermodynamic Arrow of Time}, Phys. Rev. E \textbf{81}, 061130 (2010).
\bibitem{2Law} J. Gemmer, A. Otte, and G. Mahler, \textit{Quantum approach to a derivation of the second law of thermodynamics}, Phys. Rev. Lett. \textbf{86}, 1927 (2001).
\bibitem{thermal1} J.\,M. Deutsch, \textit{Thermodynamic entropy of a many-body energy eigenstate}, New J. Phys. \textbf{12}, 075021 (2010).
\bibitem{thermal2} A. Dymarsky, N. Lashkari, and H. Liu, \textit{Subsystem eigenstate thermalization hypothesis}, Phys. Rev. E \textbf{97}, 012140 (2018).
\bibitem{thermal3} L.\,Kh. Joshi, A. Elben, A. Vikram, B. Vermersch, V. Galitski, and P. Zoller, \textit{Probing many-body quantum chaos with quantum simulators},
Phys. Rev. X \textbf{12}, 011018 (2022).
\bibitem{NonMark1} H. Spohn, \textit{Kinetic equations from Hamiltonian dynamics: Markovian limits}, Rev. Mod. Phys. \textbf{52}, 569 (1980).
\bibitem{NonMark2} M. Merkli, \textit{Quantum Markovian master equations: Resonance theory shows validity for all time scales}, Ann. Phys. \textbf{412}, 167996 (2020).
\bibitem{NonMark3} D. Chruscinski, \textit{Dynamical maps beyond Markovian regime}, Phys. Reports \textbf{992}, 1 (2022).
\bibitem{NonMark4} M. Xu, V. Vadimov, J.\,T. Stockburger, and J. Ankerhold \textit{Colloquium: Simulating non-Markovian dynamics in open quantum systems}, Rev. Mod. Phys. - Accepted (2026).
\bibitem{Interface} A.\,P. Meilakhs, \textit{Transmission of waves and particles through the interface: Reversibility and coherence}, Ann. Phys. \textbf{466}, 169686 (2024).
\bibitem{Larotonda} A.\,P. Meilakhs, C. Pastorino, M. Larotonda, \textit{Experimental demonstration of an analogy between optical non-coherence and irreversibility of heat transport},  arXiv preprint arXiv:2505.09522 (2025).
\bibitem{Optics} M. Born, E. Wolf, \textit{Principles of Optics} (Pergamon, Oxford, 1980)  {[}Chapter VII, \S 5; Chapter X, \S\S 1-4 {]}.
\bibitem{Knight}  C.\,C. Gerry, P. Knight \textit{Introductory quantum optics} (Cambridge University Press, New York, 2005).
\bibitem{Entanglement} K.\,C. Tan and H. Jeong, \textit{Entanglement as the symmetric portion of correlated coherence}, Phys. Rev. Lett. \textbf{121}, 220401 (2018).
\bibitem{NaiveLie} J. Stillwell, \textit{Naive Lie Theory} (Springer, New York, 2008).
\bibitem{MarkovBook} J.\,R. Norris,  \textit{Markov Chains} (Cambridge University Press, Cambridge, 1997).
\bibitem{LdQuant} L.\,D. Landau and E.\,M. Lifshitz, \textit{Course of Theoretical Physics, Vol. 3: Quantum mechanics} (Butterworth-Heinemann, Oxford , 1981) {[}Chapter VI, \S 40{]}.
\bibitem{NoJump} H.\,D. Zeh, \textit{There are no quantum jumps, nor are there particles!}, Phys. Lett. A \textbf{172}, 189-192 (1993).
\bibitem{LdKin} L.\,D. Landau and E.\,M. Lifshitz, \textit{Course
of Theoretical Physics, Vol. 10: Physical Kinetics} (Pergamon, New
York, 1981) {[} Chapter XI, \S 79; Chapter VII, \S 67{]}.
\bibitem{NanoWires1} G. Badawy, E.\,P.\,A.\,M. Bakkers, \textit{Electronic Transport and Quantum Phenomena in Nanowires}, Chem. Rev. \textbf{124} 2419 (2024).
\bibitem{NanoWires2} D. Shiri, R. Nekovei, A. Verma,  \textit{Low-temperature electron transport in [110] and [100] silicon nanowires: a DFT-Monte Carlo study},  Front. Nanotechnol \textbf{6}, 1494814 (2024).
\bibitem{Graphene} M.\,J.\,J. Mangnus, F.\,R. Fischer, M.\,F. Crommie, I. Swart, and P.\,H. Jacobse, \textit{Charge transport in topological graphene nanoribbons and nanoribbon heterostructures}, Phys. Rev. B \textbf{105}, 115424 (2022).
\bibitem{NanoCoherent1} J. Chen, G. Zhang, B. Li, \textit{Phonon coherent resonance and its effect on thermal transport in core-shell nanowires}, J. Chem Phys \textbf{135}, 104508 (2011).
\bibitem{NanoCoherent2} J. Chen, G. Chen and Z. Wang, \textit{Thermal transport and phonon localization in periodic h-GaN/h-AlN superlattices}, J. Phys.: Condens. Matter \textbf{36}, 045001 (2024).

\end{thebibliography}
\end{document}